# Accurate Vertical Ionization Energy and Work Function Determinations of Liquid Water and Aqueous Solutions

Stephan Thürmer,[1]* Sebastian Malerz,[2] Florian Trinter,[2,3] Uwe Hergenhahn,[2] Chin Lee,[2,4,5] Daniel M. Neumark,[4,5] Gerard Meijer,[2] Bernd Winter,[2]* and Iain Wilkinson[6]*

[1] *Department of Chemistry, Graduate School of Science, Kyoto University, Kitashirakawa-Oiwakecho, Sakyo-Ku, Kyoto 606-8502, Japan*
[2] *Molecular Physics Department, Fritz-Haber-Institut der Max-Planck-Gesellschaft, Faradayweg 4-6, 14195 Berlin, Germany*
[3] *Institut für Kernphysik, Goethe-Universität, Max-von-Laue-Straße 1, 60438 Frankfurt am Main, Germany*
[4] *Chemical Sciences Division, Lawrence Berkeley National Laboratory, Berkeley, CA, 94720 USA*
[5] *Department of Chemistry, University of California, Berkeley, CA, 94720 USA*
[6] *Department of Locally-Sensitive & Time-Resolved Spectroscopy, Helmholtz-Zentrum Berlin für Materialien und Energie, Hahn-Meitner-Platz 1, 14109 Berlin, Germany*

**ORCID**
*ST: 0000-0002-8146-4573*
*SM: 0000-0001-9570-3494*
*FT: 0000-0002-0891-9180*
*UH: 0000-0003-3396-4511*
*CL: 0000-0001-9011-0526*
*DMN: 0000-0002-3762-9473*
*GM: 0000-0001-9669-8340*
*BW: 0000-0002-5597-8888*
*IW: 0000-0001-9561-5056*

*Corresponding authors: thuermer@kuchem.kyoto-u.ac.jp; winter@fhi-berlin.mpg.de; iain.wilkinson@helmholtz-berlin.de*

# Abstract

The absolute-scale electronic energetics of liquid water and aqueous solutions, both in the bulk and at associated interfaces, are the central determiners of water-based chemistry. However, such information is generally experimentally inaccessible. Here we demonstrate that a refined implementation of the liquid microjet photoelectron spectroscopy (PES) technique can be adopted to address this. Implementing concepts from condensed matter physics, we establish novel all-liquid-phase vacuum and equilibrated solution-metal-electrode Fermi level referencing procedures. This enables the precise and accurate determination of previously elusive water solvent and solute vertical ionization energies, VIEs. Notably, this includes quantification of solute-induced perturbations of water's electronic energetics and VIE definition on an absolute and universal chemical potential scale. Defining and applying these procedures over a broad range of ionization energies, we accurately and respectively determine the VIE and oxidative stability of liquid water as $11.33 \pm 0.02$ eV and $6.60 \pm 0.08$ eV with respect to its liquid-vacuum-interface potential and Fermi level. Combining our referencing schemes, we accurately determine the work function of liquid water as $4.73 \pm 0.09$ eV. Further, applying our novel approach to a pair of exemplary aqueous solutions, we extract absolute VIEs of aqueous iodide anions, reaffirm the robustness of liquid water's electronic structure to high bulk salt concentrations (2 M sodium iodide), and quantify reference-level dependent reductions of water's VIE and a $0.48 \pm 0.13$ eV contraction of the solution's work function upon partial hydration of a known surfactant (0.025 M tetrabutylammonium iodide). Our combined experimental accomplishments mark a major advance in our ability to quantify electronic-structure interactions and chemical reactivity in liquid water, which now explicitly extends to the measurement of absolute-scale bulk and interfacial solution energetics, including those of relevance to aqueous electrochemical processes.

# Introduction

Knowledge of the electronic structure of liquid water is a prerequisite to understand how water molecules interact with each other and with dissolved solutes in aqueous solution. Here, the valence electrons play a key role because their energetics govern chemical reactions.[1] One quantity of particular interest is water's lowest vertical ionization energy, VIE (or equivalently vertical binding energy, VBE), which is a measure of the propensity to detach an electron under equilibrium conditions and thus determines chemical reactivity.[2] More precisely, $VIE_{vac}$ (where the 'vac' subscript refers to energetic referencing with respect to vacuum) is the most probable energy associated with vertical promotion of an electron into the vacuum, *i.e.*, without giving it any excess energy, and with no nuclear rearrangement being involved. Such $VIE_{vac}$ values are most readily accessed using photoelectron spectroscopy (PES) – usually from gases, molecular liquids, or molecular solids – and are identified as the maximum intensities of (primary, directly-produced) photoelectron peaks.

Generally, in the condensed phase, PES features cannot be correlated with isolated molecular states, but are instead considered (particularly in crystalline samples) to arise from band structures, dense collections of states born from extended inter-atomic interactions.[N1] Broad PES features are most often observed, from which it is often impossible to reliably extract valence VIE values. However, in molecular liquids and molecular solids, peak structures usually remain isolable, with associated $VIE_{vac}$ values regularly being extracted and described within a molecular physics framework. Here, simple molecular orbital formalisms are adopted, with the peak structures ascribed to liberation of electrons from specific orbitals. Adopting such an approach, the molecular orbitals of the water monomer have been considered to be only weakly perturbed by hydrogen bonding in the liquid phase, without specific regard for inter-monomer interactions or explicit consideration of the aqueous interface. The lowest $VIE_{vac}$ value of water has correspondingly been assigned to ionization of the non-bonding $1b_1$ highest occupied molecular orbital (HOMO) in the gas,[3] liquid,[4] and solid[5] phases. In fact, this molecular electronic structure description, and a vacuum level energy referencing approach, has almost exclusively been adopted in the interpretation of liquid-phase PES spectra.[2, 6, 7] This is in spite of liquid water (and aqueous solutions) exhibiting both molecular [4, 8-11] and dispersed 'band'[7, 8, 12-17] electronic structure signatures. Naturally, this raises the questions of how liquid water should be placed between the aforementioned molecular and condensed matter conceptual frameworks, and specifically what can be learned by applying concepts from the latter to the PES of liquid water and aqueous solutions.

Within a condensed-matter framework and at thermodynamic equilibrium, the available states (or bands) of a system, are separated into occupied and unoccupied components around the Fermi level, $E_F$. As a precisely defined thermodynamic quantity, energy referencing with respect to $E_F$ engenders direct comparison of system energetics between condensed-phase samples and the ready relation of those energetics to additional thermodynamic quantities. Such a useful energetic reference is readily accessible in metals using PES, where $E_F$ lies within the available states and defines the upper electronic occupation level. In contrast, in semi-conductors, $E_F$ is placed within a 'forbidden' band gap (devoid of states) and is thus, directly at least, inaccessible using the PES technique; $E_F$ is notably not an electronic state that can donate or accept electrons here, rather it corresponds to a thermodynamic energy level. Liquid water, like most other liquids, can be classified as a wide-band-gap semiconductor,[18-20] with a generally inaccessible Fermi level. Upon first consideration, liquid water may, therefore, seem unsuited to an $E_F$ energy referencing scheme. Clearly, the solid-state custom of indirectly energy-referencing semi-conductor PES spectra to $E_F$ via metallic reference sample is much more difficult to apply to volatile and potentially charged aqueous-phase samples.

The $VIE_{vac}$ values predominantly considered in liquid-phase PES experiments so far, as well as any VIE values determined with respect to $E_F$, $VIE_{EF}$, arise from the cumulative energetics of a photoemission process. This includes the effects of collective phenomena (hydrogen bonding, inhomogeneous broadening etc.), electron transport, and an interface (typically liquid-vacuum),[21-23] where the latter has yet to be explicitly addressed in liquid-phase PES studies. In liquid water, the ionization energies are specifically affected by inhomogeneous and fluxional intermolecular interactions (*i.e.,* hydrogen bonding). Here, the associated energetics vary over the transition region spanning the aqueous bulk and the liquid interface through which photoelectrons must traverse to escape into vacuum. These properties are closely related to distinctive condensed-matter system descriptors that are of particular relevance to photoemission, such as electrical conductivity, chemical potential ($\mu$, equivalent to $E_F$), electrochemical potential ($\bar{\mu}$), work function ($e\Phi$), surface dipole, and surface (dipole) potential ($\chi^d$ or $e\varphi_{outer}$).[24-26] We present an overview of the relations between these parameters, with a focus on the liquid water system, in Figure SI-1 of the supporting information (SI) and note that even after many years of aqueous-phase PES research, previous evaluations of liquid water's (lowest) $VIE_{vac}$ values[4, 27-29] have barely considered these condensed matter descriptors. In other words, more differential probes of the bulk and interfacial electronic structure properties of liquid water and aqueous solutions have barely been addressed in PES experiments.[N2]

We show here that application of concepts from condensed-matter physics to liquid-jet (LJ) PES enables a significant expansion of our understanding of the electronic structure of liquid water. Towards that wider goal we pronounce two immediate aims. The first is to determine an accurate value of the lowest vacuum-level-referenced VIE of liquid water, $VIE_{vac,1b1(l)}$ (equivalent to its HOMO or $1b_1$ orbital ionization energy). Perhaps surprisingly, after more than 15 years of research, the value of this quantity remains controversial, mirroring key shortcomings in previous experiments. We address these deficiencies here and identify the need for additional spectroscopic information. For this particular task, the missing quantity is the (yet-to-be-discussed, although previously alluded to [4,30, 31]) low-energy electron cutoff in the liquid-water PES spectrum, a commonly measured parameter in solid-state PES.[23, 32-35] Motivated by a possible depth dependence of $VIE_{vac,1b1(l)}$ (*i.e.*, of neat water), we utilize the cutoff spectral feature to report the first systematic study of this quantity over a large range of photon energies, spanning the (vacuum) ionization threshold region up to more than 900 eV above it. We apply the same concepts to determine water's lowest VIE from exemplary aqueous solutions, $VIE_{vac,1b1(sol)}$, in addition, *i.e.,* detecting the solute-induced effect on water's electronic structure. We similarly demonstrate how to extract the VIEs of aqueous solutes, $VIE_{vac,solute}$, over a broad range of concentrations. Our second principal objective is to demonstrate how to measure $E_F$ and $e\Phi$ of liquid water and aqueous solutions. We will discuss the meaning and importance of $E_F$ in the case of the liquid water system, with the main goal of obtaining liquid-phase VIEs referenced to its Fermi level ($VIE_{EF}$), including those of neat water ($VIE_{EF,1b1}$), the aqueous solvent ($VIE_{EF,1b1(sol)}$), and associated solutes ($VIE_{EF,solute}$). The successful implementation of this alternative aqueous-phase PES energy referencing scheme permits a direct comparison between liquid- and solid-phase PES results. It further enables (more) direct derivation of additional thermodynamic quantities from aqueous-phase VIE measurements, including redox energetics. The combination of the $VIE_{EF}$ information with respective $VIE_{vac}$ measurement results allows $e\Phi$ values to be derived and the explicit characterization and quantification of aqueous interfacial effects. Finally, we evaluate the challenges in characterizing Fermi level alignment between solutions and reference metals based on the currently available experimental methods, as we start to bridge the gap between aqueous-phase and solid-phase PES.

## LJ-PES from water and aqueous solution

### The common experimental approach

We begin with short overviews of the LJ-PES technique, the commonly adopted LJ-PES vacuum energy referencing method, and the current challenges in measuring accurate $VIE_{vac}$ values of liquid water and solutions more generally. We also present some useful considerations on the application of a VIE scale to condensed-phase PE spectra in SI Section 1, which we apply from here onwards. Since the experimental breakthrough in detecting photoelectron spectra from aqueous solutions, marked by the availability of vacuum liquid microjets[36, 37] over 20 years ago, a flurry of LJ-PES measurements has been conducted. Such measurements have greatly advanced our understanding of the electronic structure of aqueous solutions, in the bulk and at the solution–vacuum interface, as has recently been reviewed.[38] Notably, however, aside from very few exceptions, previous LJ-PES measurements have garnered the bare minimum spectral information, for which it has sufficed to detect a narrow range of electron kinetic energies, eKEs, of the emitted photoelectron distributions. For example, from aqueous LJs and their evaporating vapor layer, the characteristic eKEs of a solute or liquid water ionization feature of interest, $VIE_{vac,(l)}$, and the lowest energy gas-phase ionization peak, $VIE_{vac,1b1(g)}$, can be simultaneously determined. The latter value is accurately known ($12.621 \pm 0.008$ eV[3]), and from the difference of the measured peak positions, $\Delta E_{g-l} = VIE_{vac,1b1(g)} - VIE_{vac,1b1(l)}$, $VIE_{vac,1b1(l)}$ can (in principle) be determined.[4, 28, 36] Adopting this procedure, here referred to as Method 1, vacuum-level energy referencing and production of the aqueous photoemission spectrum is achieved without need for further information. This simple and highly convenient molecular-physics approach, which is however challenging to accurately apply, as we will show below, is illustrated in Figure 1A. There, we depict the measured valence photoemission spectrum of liquid water, *i.e.*, the kinetic energy distribution curve of the emitted photoelectrons, and the energy difference, $\Delta E_{g\text{-}l}$, between the lowest energy liquid-, $1b_{1(l)}$, and gas-phase, $1b_{1(g)}$, water ionization features.

LJ-PES experiments commonly use rather high photon energies, typically some tens or more electron volts above the relevant ionization thresholds. Such photon energies sufficiently separate directly-produced photoelectron peaks from the low-energy background of inelastically scattered electrons[23] and minimize scattering-induced distortions of the PE peaks themselves[30] (owing to the fact that electron scattering is almost exclusively governed by electronic excitations at such photon and kinetic energies[39]). The vast majority of LJ-PES studies have adopted such photon energies to establish solute *core*-level energies, with the measured chemical shifts serving as a reporter of changes in the chemical environment. Small discrepancies in absolute core-level energies among different laboratories typically have little consequence on the main observations and derived statements. Similarly, the large body of studies of Auger decay and other autoionization processes from the aqueous phase[40-43] would be barely affected by small uncertainties in absolute electron energies. This is in contrast to the situation with valence LJ-PES, which has been far less explored[2, 44, 45] despite the primary importance of the lowest-ionization energies in driving aqueous-phase chemistry.[2] In this case, after more than 15 years of active high-energy-resolution LJ-PES research[38, 43] (and with concomitant advancement of aqueous electronic structure calculations and spectral simulation methods[8, 9, 41, 46-52]), an experimental advance and alternative terminology must be adopted to enable unequivocal and accurate valence VIE determinations with respect to the vacuum level. Related developments are needed to permit $E_F$ (or system chemical potential) energy referencing of LJ-PES spectra, robust $e\Phi$ extractions from liquid samples, and direct comparisons of liquid- and solid-phase absolute-scale electronic energetics.

To understand the shortcoming of previous studies it is sufficient to discuss why the exact value of $\text{VIE}_{\text{vac},1b1(l)}$ from neat water continues to be debated, spanning a 0.5 eV range between $11.16 \pm 0.04$ eV[4] and $11.67 \pm 0.15$ eV[29]. All previously reported reference values were obtained using Method 1, from a mere $\Delta E_{g\text{-}l}$ measurement which neither requires the determination of absolute eKEs nor an exact calibration of the applied photon energy. However, a seemingly simple measurement of $\Delta E_{g\text{-}l}$ is difficult to accomplish due to the multiple sample charging effects and contact-potential differences that occur in LJ spectrometer systems (see the discussion in Section 2 in the SI and Refs. [7, 28, 53, 54]). Accurate $\Delta E_{g\text{-}l}$ measurements are further complicated by the temporal variation of surface potentials within LJ-PES apparatuses, due to the continuous evaporation of LJs and the establishment of stable, adsorbed surface layers within spectrometers. All of these perturbing influences generate electric fields between the sample and the electron detector, which affect the photoelectrons from the gas and liquid phases differently and have to be precisely accounted for to achieve the 'true' (*i.e.*, undisturbed) $\Delta E_{g-l}$ value. As knowledge about the relevant effects and methods for their elimination continues to evolve,[28, 53, 54] reported $\Delta E_{g-l}$ values, and thus deduced $VIE_{vac,1b1(l)}$ values continue to vary from laboratory to laboratory, which explains the scatter of the reported energies mentioned above.

Efforts to measure accurate $\Delta E_{g-l}$ values center around the minimization or even elimination of the effects of perturbing potentials, compensating (electrokinetic and other forms of) charging of the LJ and other local potentials to achieve what we refer to as 'field-free' conditions. The primarily adopted method achieves this by implementing a small but precisely determined salt concentration in water at a given solution flow rate and temperature.[28] Alternatively, the provision of field-free conditions through application of a compensating bias voltage to a LJ has been discussed.[29, 31] In spite of such compensation efforts, the stabilization of spectrometer potentials occurs on the order of tens of minutes to hours after LJs are started or experimental parameters are adjusted, for example, when cold trap coolant is replenished. As we show in Figure SI-2, the apparatus potentials change dramatically (more than 100 meV) over time upon introducing water vapor into the experimental vacuum chamber, while eventually settling into an equilibrium. Unsurprisingly, these effects are difficult to quantify for a given experimental setup and operational conditions.

Here we highlight another potentially crucial and barely realized issue with Method 1, namely the meaning of the vacuum level. We have introduced $VIE_{vac}$ above without providing a sufficiently accurate definition of the relevant vacuum level in a LJ-PES experiment. $VIE_{vac,1b1(g)}$ (like any other gas-phase ionization energy) is necessarily referenced to the vacuum level at infinity, $E_v^\infty$ (used in Figure 1A), and corresponds to the potential energy of the photoelectron at rest and at infinite separation from the photoionized sample.[24] In all previous LJ experiments, it has been implied that this same vacuum level is applicable and accessible upon ionization of liquid water, with existing $VIE_{vac,1b1(l)}$ values being consequently referenced to $E_v^\infty$ via $VIE_{vac,1b1(g)}$. Adopting this assumption, the most probable (vertical) gas- and liquid-phase ionization energies have been taken as the maxima of the gas- and liquid-phase photoelectron (PE) peak fits within an encompassing spectrum. The consequences of this assumption will be further discussed below.

A yet further encountered and momentous oversight in previous LJ-PES studies is the determination of aqueous-phase solute $VIE_{vac}$ values ($VIE_{vac,solute}$) with reference to predetermined $VIE_{vac,1b1(l)}$ values measured from neat water, ideally under field-free conditions. That is, in (almost) all previous LJ-PES valence and a number of core-level studies spanning a broad range of aqueous solutions,[2, 38] the $VIE_{vac,1b1(l)}$ value (*i.e.*, from neat water) has in fact been used (as is) to calibrate $VIE_{vac,solute}$ values. Specifically, the energy difference between the solute PE peak position and lowest-energy solvent PE peak position, $VIE_{vac,1b1(sol)}$, has been used, under the generally erroneous assumption that $VIE_{vac,1b1(sol)} = VIE_{vac,1b1(l)}$. This is illustrated in the inset of Figure 1A, where $\Delta E_{l-l}$ is the measured energy difference between two liquid-phase peaks, the lowest ionization energy, $1b_{1(l)}$, solvent peak and a solute peak. This energy referencing is generally rendered meaningless when non-negligible solvent-solute interactions and/or solute-induced interfacial electronic structure changes occur. In core-level studies, often the O 1s core-level energy (established for neat water only, again under field-free conditions)[55] has alternatively been used to similarly energy-reference $VIE_{vac,solute}$ values, with the same fundamental deficiencies. Such practices imply that solute-induced water electronic structure and solution $e\Phi$ changes do not occur, an assumption which has no rigorous foundation and may easily lead to quantitative failure of this extended implementation of Method 1, as recently discussed in Ref. [7] and enunciated in Ref. [31].

Alternatively, but equally problematic, one could strive for the determination of $VIE_{vac,solute}$ with reference to $VIE_{vac,1b1(g)}$, using the basic variant of Method 1 (*i.e.*, the hypothetical field-free variant of what is shown in the main section of Figure 1A). Yet, as detailed above, only if the region between the LJ interface and detector were field-free, could the measured electron energies from the gas-phase molecules be directly related to those from the liquid phase. For almost all solutions, field-free conditions are not or cannot be established in the experiment, and the same problems remain as for neat water. Thus, any additional field introduced to the solution – via electrokinetic charging, ionization-induced charging, or surface dipoles – renders the direct $\Delta E_{g-l}(solute)$ energy referencing via (extrinsically field-free) values of $VIE_{vac,1b1(g)}$ questionable. With Method 1, the relative contributions to the sample charging cannot be quantified, and field-free conditions thus only arise from the fortunate mutual compensation of any charging and/or differential $e\Phi$ effects.

Furthermore, and more fundamentally, the effects of any intrinsic and non-negligible interfacial dipole potential, $\chi^d$, at the water liquid-vapor-phase interface[56] could lead to intrinsic offsets of $\Delta E_{g-l}$ from its true value, potentially compromising energy referencing Method 1. The value of the liquid water interfacial surface dipole potential has yet to be directly experimentally determined, although it has been inferred to amount to a few tens of meV in neat water,[57,58] with associated theoretical predictions[56,59-62] of $\chi^d$ varying significantly. In aqueous solutions, the value of $\chi^d$ is expected to be highly solute- and concentration-dependent,[56] calling the extended Method 1 energy referencing schemes for aqueous solutions further into question. Hence, to uniquely and generally interrogate both solute and solvent electronic structure on an absolute energy scale, a novel and robust experimental procedure that relies on an energy reference other than $VIE_{vac,1b1(g)}$ must be developed.

**Condensed-matter approach and absolute energy reference**

Above we have seen that an approximate value of $VIE_{vac,1b1(l)}$ from neat liquid water – with up to 0.5 eV uncertainty, depending on the care taken to compensate extrinsic potentials – can be obtained with the conceptually simple Method 1 (Figure 1A). Adopting a more robust, absolute energy referencing method afforded using the low-energy photoelectron signal cutoff, $E_{cut}$, as widely applied in solid-state PES,[23,32-35] the field-free requirement for accurate $VIE_{vac}$ measurements is lifted. We now consider the associated energy-level diagram shown in Figure 1B to illustrate this more robust and generally applicable experimental approach. In fact, as a requirement for an accurate $VIE_{vac,1b1(l)}$ (or alternative liquid-phase $VIE_{vac}$) determination, a negative bias voltage should be deliberately applied between the LJ and the electron analyzer (orifice), imparting a well-defined additional eKE to the liquid-phase photoelectrons via an accelerating field, $E_{acc}$ (indicated as black dotted line in Figure 1B); we explain why the application of a bias voltage is indispensable below. Hence, a prerequisite for this approach is a sufficiently electrically conductive sample (that supports the applied bias), held in direct electrical contact with the electron analyzer via a (stable) DC power supply. Not only does this allow the unequivocal resolution of the true value of $VIE_{vac,1b1(l)}$ from neat water, the respective value (as well as any associated solute $VIE_{vac,solute}$) can also be accessed from *any* aqueous solution. In fact, the same methodology is also directly applicable more generally, for example, to organic solutions. Moreover, novel information on the solution–vacuum interface is conveniently provided.

The full LJ-PES spectrum from neat liquid water is sketched in Figure 1B. The case of a photon energy sufficiently in excess of the first three ionizing transition thresholds of liquid water ($1b_1^{-1}$, $3a_1^{-1}$, and $1b_2^{-1}$ in a molecular-physics description) to yield undistorted primary photoelectron peaks is illustrated. Spectra associated with grounded (grey curve) and negatively biased (blue curve) liquid samples are shown. In the biased case, the entire liquid-phase spectrum experiences a rigid energy shift, equivalent to the (negative) bias voltage (see Figure SI-3 for an experimental example of this effect). The exact value of the bias voltage is rather irrelevant for the present purpose. Unlike in Figure 1A, the spectra in Figure 1B encompass the low-KE tail, LET, which terminates the spectrum at eKE = 0 eV.[N3] The LET comprises electrons which have lost most of their energy due to various inelastic scattering processes, and have just enough energy to overcome the surface barrier of the sample. They are accordingly expelled with quasi-zero kinetic energy, signified here by the $E_{cut}$ label, with $E_{cut}$ defining the energetic zero from the perspective of a photoelectron leaving the sample[35]. Hence, the concurrent measurement of $E_{cut}$ (=0 eV) and the $VIE_{vac}$ values of interest – such as $VIE_{vac,1b1(l)}$ – allows the unique and self-consistent assignment of an eKE reference to the LJ-PES data, irrespective of *any* perturbing potentials, intrinsic or extrinsic. From Figure 1B, it is seen that the eKE of the $1b_{1(l)}$ peak can be accurately determined via its energy separation from $E_{cut}$, *i.e.*, the spectral width, $\Delta E_w$. The associated VIE is correspondingly determined as $VIE_{vac,1b1(l)}$ = hν - $KE_{1b1(l)}$, where $E_{cut}$ is set to 0 eV and it is implied that the photon energy is precisely known (we discuss procedures to precisely determine hν for various light sources in the SI Section 3). This procedure of measuring the *full* PES spectrum (or, at least, the LET region and the PE features of interest under the same conditions) will be referred to as Method 2 in the following. Importantly, gas-phase peaks or referral to $VIE_{vac,1b1(g)}$ are now irrelevant for the accurate extraction of $VIE_{vac,1b1(l)}$ (or any other solvent or solute VIE). Furthermore, a favorable side effect of applying a high enough bias voltage is that the liquid-phase PE spectrum can be obtained (essentially) free from otherwise overlapping gas-phase signal, as is indicated by the missing sharp $1b_{1(g)}$ peak in the blue curve in Figure 1B. In that case, the varying electrostatic potential between the biased liquid sample and the grounded electron analyzer results in a gas-phase peak broadening and a differential

gas–liquid shift which is sufficient to move the gas-phase peak centers out of the liquid phase spectrum. Thus, the gas-phase features can almost be fully pushed out of the spectral range of interest. Notably, however, it is impossible to fully suppress the gas-phase signal at the energy position of the liquid spectrum by applying a bias, as some gas-phase molecules will always reside directly above the surface and experience the full bias potential.

We have not yet thoroughly motivated the rationale for conducting experiments on a negatively biased sample, which so far was rarely practiced in liquid-phase PES. In the case of an unbiased LJ, the spectrum of the LET is typically obscured by the measurement process, as the PE distribution is modified by additional scattering inside the electron analyzer and then generally arbitrarily terminated at a low-energy cutoff, $E_{cut(A)}$, by the analyzer's own internal work function.[34, 35] This makes an accurate distinction of the true sample cutoff, $E_{cut(s)}$, impossible. The overlapping cutoffs for the unbiased liquid-water jet are correspondingly depicted in the bottom part and grey spectrum in Figure 1B, with this spectrum being energetically-aligned with that shown in Figure 1A. As partially highlighted in Figure 1B, only by applying a sufficiently large negative bias voltage to the liquid jet can the LET curve of the sample and the secondary electron signals produced in the analyzer be well separated, the arbitrary $E_{cut(A)}$ threshold be far exceeded, and $E_{cut(s)}$ be precisely determined.

Thus far we did not comment on the appropriate vacuum reference level for Method 2. As alluded to above, gas-phase and condensed-phase PES measurements in principle refer to different vacuum levels. This is connected to the presence of a surface, through which the photoelectrons have to traverse as the final step in a condensed-phase PE process.[23] $E_{cut}$ marks the minimum energy for a photoelectron to surmount the surface barrier and be placed at rest at a point in free space just outside the surface, overcoming $e\Phi$ (*i.e.*, where the electron image potential at the surface drops to zero and at a distance from the surface that is much smaller than the dimensions of the surface itself).[24] This connects all energies inferred with Method 2 to the *local* vacuum level, $E_v^{loc}$, but not necessarily to $E_v^\infty$. In aqueous solutions, the offset of $E_v^{loc}$ with respect to $E_v^\infty$ can be related to the outer (Volta) potential $e\varphi_{outer}$ or $\chi^d$,[59] note the small $E_v^{loc}$ versus $E_v^\infty$ difference labeled $\chi^d$ / $e\varphi_{outer}$ in Figs. 1A and 1B, where the panels connect. Generally, an intrinsic millivolt to volt scale dipolar surface potential, $\chi^d$, is expected to occur at the aqueous liquid-gas interface as the molecular density and hydrogen bond structure of bulk liquid water or an aqueous solution evolves from fully hydrated to partially hydrated and to increasingly isolated molecules in the gas phase. A range of experimental[57, 58, 63] and theoretical[56, 59-62] studies have been performed to infer or calculate the net dipolar alignment and associated interfacial potential difference in the neat (or nearly neat) water case. While few tens of meV values have been inferred experimentally,[57, 58] a consensus on the value of $\chi^d$ at the water liquid-vapor-phase interface has yet to be reached from a theoretical perspective, and direct experimental measurements have not, to our knowledge, been reported. Relating this to the present discussion, $\chi^d$ clearly only emerges within a condensed-matter description of the aqueous-phase electronic structure. Furthermore, any non-negligible $\chi^d$ value would differentially affect electrons born at different points across the aqueous bulk to gas-phase transition region. Correspondingly, energy referencing Method 2 and the thus far adopted direct $\Delta E_{g\text{-}l}$ energy referencing approach, Method 1, can be expected to yield inherently different $VIE_{vac,1b1(l)}$ values if a significant liquid water $\chi^d$ pertains.

Moving beyond our primary consideration of neat liquid water, Method 2 can also be applied without amendment to aqueous (or other) solutions, as shown in Figure 1C. We can thus determine $VIE_{vac,1b1(sol)}$ with the same high accuracy as $VIE_{vac,1b1(l)}$ for neat liquid water, with the additional possibility of precisely determining other aqueous-phase solvent and solute VIEs. $VIE_{vac,1b1(l)}$, $VIE_{vac,1b1(sol)}$ and $VIE_{vac,solute}$ are again obtained as $VIE_{(l)} = h\nu - KE$ with $E_{cut}$ defining zero KE. A solute-induced change of the former is seen to directly correspond to a change in the measured $1b_1$ ionization feature KE, corresponding to the different values of $\Delta E_w$ and $\Delta E_{w(sol)}$. We show an additional high-KE peak in Figure 1C to exemplarily illustrate the photoionization of a solute component. We emphasize that in the presence of a solute, surface potentials (in addition to the aforementioned extrinsic fields) are likely to be modified, generally making it impossible to establish the field-free conditions needed to directly apply Method 1. Its extended variant – measurement of $\Delta E_{l\text{-}l}$ and energy referencing to the field-free value of $VIE_{vac,1b1(l)}$ – as has so far been utilized to obtain reference energies for $VIE_{vac,solute}$ values, is similarly invalidated. Method 2, on the other hand, is not affected and thus permits direct access to absolute VIE changes between aqueous (or alternative) solutions for the first time. We further stress that Method 2 probes VIEs with respect to the local vacuum level $E_v^{loc}$ and that the energetic position of $E_v^{loc}$ with respect to $E_v^\infty$ generally varies depending on the solution (note that the schematic biased spectra in Fig 1C are arbitrarily aligned to the low-energy cutoff, which simultaneously aligns $E_v^{loc}$). Analogous to Figure 1B, we illustrate the spectrum measured from an unbiased aqueous solution at the bottom of Figure 1C, which highlights the overlapping sample and spectrometer LET curves and depicts the general inaccuracy of unbiased $\Delta E_{l\text{-}l}$ measurements when energy referenced using previously determined field-free (neat water) $VIE_{vac,1b1(l)}$ values (as shown in the inset of Figure 1A).

**Fermi-level referencing and solution work functions**

In the following we consider additional steps beyond the absolute, vacuum level energy referencing ability of Method 2 (Figs. 1B and 1C) and address the interfacial electronic structure information that becomes accessible using a condensed-matter framework and associated experimental approach. This leads us to attempt to determine $E_F$ and $e\Phi$ in both water and aqueous solutions, with the latter providing a means to differentiate between solute-induced changes of (bulk or surface) liquid electronic structure or interfacial effects. Correspondingly, we briefly explain the concepts of $E_F$ and $e\Phi$. $E_F$ is formally equivalent to the chemical potential, $\mu$, and at thermodynamic equilibrium is the energy at which a (potentially hypothetical) electronic state has 50% probability of being occupied at fixed temperature and any given time. The position of $E_F$ throughout matter in electrical equilibrium assumes the same thermodynamic value. This makes $E_F$ an advantageous energy reference in condensed-matter spectroscopies, especially for metallic samples, in which electrons occupy states up to $E_F$, and which can be directly measured using photoemission. $E_F$ is conceptually connected to two additional important quantities, the electrochemical potential, $\bar{\mu}$, and the work function, $e\Phi$. $\bar{\mu}$ is the energy required to bring an electron at rest *at* infinity into the bulk of the material. Hence, the sum of $E_F$ and $\bar{\mu}$ is equivalent to $E_v^\infty$ (and in a metal, the energy of $\bar{\mu}$ with respect to $E_F$ is equivalent to the electron affinity). In contrast, $e\Phi$ is the minimum energy required to remove an electron at $E_F$, deep inside the material, and place it at rest at a point in free space just outside the surface, thus connecting to the local vacuum level, $E_v^{loc}$. $E_v^{loc}$ and $e\Phi$ are correspondingly local properties of a surface which can change widely depending on the surface conditions.[N4]

Figure 1D depicts the energetic alignment of $E_F$ for (grounded) liquid water and a (grounded) metal, which implies electrical contact between the liquid, the metal sample, and the analyzer. The exact meaning of 'aligning the Fermi level' of a solid and a liquid will be detailed in the Discussion section. To generate accurate PES results, sufficient electrical conductivity must be engineered between all of these elements while suppressing parasitic extrinsic potentials, such as the aforementioned LJ streaming potential. In these conditions, $E_F$ can be directly measured from a metal, as indicated by the red archetypal spectrum on the right of Figure 1D. The water sample, which is in direct electrical contact with the metallic reference sample and the analyzer, is then separately probed under the same conditions to produce the blue water spectrum on the left of Figure 1D (identical to that shown in Figure 1A). Sequential PES measurements from these two samples accordingly provides a means to formally assign $E_F$ to liquid water (as implied in Figure 1D), and hence defines the energy scale needed to determine water's ionization energy with respect to the Fermi level, $VIE_{EF,1b1(l)}$.[N5] Such pairwise measurements will be reported here, where extensive efforts have been made to measure the LJ sample and metal reference spectra under as similar conditions as possible, for example by recording the latter in the presence of the LJ in operation to capture any potentially distorting influences of the LJ. The measured $E_F$ position from the metal reference sample was found to remain constant within ~2 meV, regardless of conditions inside the vacuum chamber or whether the LJ was on or off. Despite this, our associated experimental approach, referred to in the following as Method 3, does however have a notable deficiency. As the electrons emitted from the metal are measured without crossing the solution – vacuum interface, any parasitic potentials and surface effects uniquely present on the LJ are not captured by Method 3. Extrinsic potentials, such as the streaming potential and light-induced surface charging, which are dependent on the solution and various experimental parameters, pose a new and unique challenge to the Fermi-referencing approach.[N6] In order to accurately and generally perform the $E_F$ referencing procedure, the electrons from the metal sample would also need to be detected following traversal of the solution – vacuum interface, for example using a PES-compatible solution-on-metal sample system incorporating a continuous solution flow (to avoid sample contamination and cumulative photo-induced degradation). With presently available experimental techniques (including electron-permeable flow cell windows[64]), such a measurement remains elusive[65] due to the small electron mean free path in water.[66, 67] This constitutes one of the major challenges in measuring PES from water-solid interfaces. However, although an ideal $E_F$ alignment and single-experiment $E_F$–referenced liquid-phase PES measurement (as suggested in Figure 1D) is not yet feasible, $E_F$ alignment can still be achieved via analysis of the two separately and carefully measured spectra, as we discuss below.

Arguably, Method 3 can be applied for Fermi level referencing of aqueous-phase PES spectra under favorable conditions, specifically where parasitic potentials are effectively suppressed. In general, this is *explicitly* a different acquisition condition to the field-free condition required for Method 1. The work functions, $e\Phi$, of the samples and the detection system usually differ, which results in a contact potential difference, $\Delta e\Phi$, between the analyzer, the metallic reference, and/or the LJ sample in the $E_F$-aligned case; this situation is sketched in the inset of Figure 1D. For the meaningful application of Method 3, one instead needs to find conditions in which 1) the solution conductivity is sufficiently high to enable alignment of $E_F$, by the exchange of charge between the solution and the grounding electrode, and 2) adequate suppression of both the streaming potential and ionization-induced sample charging is given. In this case, shifting of the liquid PE features with respect to $E_F$ in the measured spectrum can be avoided, *i.e.*, a direct relation between the liquid and the measured metallic reference spectrum can be established. Thus, after careful elimination of these influences, and the performance of two separate measurements to detect $VIE_{1b1(l)}$ or $VIE_{1b1(sol)}$ from the LJ and the Fermi edge from the reference metal sample, $E_F$ referencing is in principle established. We emphasize – analogous to the gas-phase referencing approach, Method 1 – that if *extrinsic* potentials (other than the aforementioned $\Delta e\Phi$) remain, *e.g.*, by insufficient compensation during the experiment, the liquid and metal spectra (*i.e.*, *measured* eKEs) are differentially affected, preventing a common energy referencing based on Method 3.

With $VIE_{vac,1b1(l)}$ determined via Method 2, a comparison to $VIE_{EF,1b1(l)}$ determined with Method 3 directly yields $e\Phi_{water}$ with the caveats described in Note N4. A conceptually similar procedure was previously applied by Tissot et al.[68] to extract $E_F$-referenced VIE values from static, low-vapor pressure, saturated (~6 M) NaCl and (~11 M) NaI aqueous solutions deposited on a gold substrate. There, the metallic and liquid features were referenced to each other under grounded conditions, with the associated approach further benefitting from being free from streaming potentials due to the static nature of the immobile liquid droplet. A value of $e\Phi$ was subsequently determined by biasing the sample and probing the associated isolated LET signal (see Note N3). However, organic impurities contained in the solutions and accumulated radiation-induced sample damage may have obfuscated the true value of $e\Phi$; both issues are generally negligibly small when using liquid-microjet sample-delivery methods.[37] A subsequent attempt to determine $e\Phi_{water}$ using core-level LJ-PES – from 50 mM

NaCl and 0.15 M butylamine aqueous solutions – was reported,[31] albeit based on the implementation of an inadequate procedure that relied on several questionable assumptions, as detailed in SI section 7.

More recently, we have been made aware of a study by Ramírez,[69] which, building on the two works mentioned above, reports $VIE_{1b1(l)}$ and work function measurements from KCl and Zobell[70] aqueous solutions to tune the aqueous redox potential; the reasons and implications for implementing such a redox couple is detailed below when we present our measurements of liquid water's work function. The $VIE_{vac}$, $VIE_{EF}$ and $e\Phi$ values notably differ from the values reported in the present work and are elaborated on in the Results & Discussion section as well as SI Section 7.

**Methods**

Experiments were performed at four facilities, equipped with different setups. Measurements at photon energies of ~15 eV, ~20 eV, ~25 eV, and ~30 eV were conducted at the DESIRS VUV beamline[71] of the SOLEIL synchrotron facility, Paris, using a novel LJ-PES apparatus.[72] The same LJ-PES setup was used for He I α (= 21.218 eV), He II α (= 40.814 eV), and He II β (= 48.372 eV) measurements in our laboratory at the Fritz-Haber-Institut (FHI), Berlin, and for measurements at photon energies of ~250 eV, ~400 eV, and ~950 eV at the P04 soft X-ray beamline[73] of the PETRA III synchrotron facility (Deutsches Elektronen-Synchrotron, DESY, Hamburg). Briefly, the LJ-PES apparatus is equipped with a Scienta Omicron HiPP-3 hemispherical electron analyzer (HEA), complete μ-metal shielding, and, when not operated at a synchrotron radiation source, a VUV5k monochromatized plasma-discharge lamp (He) for the laboratory experiments. Measurements at photon energies of ~123.5 eV, ~247 eV, ~401 eV, ~650 eV, and ~867.5 eV were additionally performed using the SOL³PES setup[74] at the U49-2_PGM-1 soft X-ray beamline[75] at the BESSY II synchrotron radiation facility in Berlin.

In the low-photon-energy synchrotron experiments at SOLEIL, the light was linearly polarized perpendicular to the plane of the laboratory floor, which was the plane spanned by the LJ and light propagation axes. The analyzer collected electrons in a backward scattering geometry, forming an angle of 40° to the light polarization direction. An energy resolution of better than 3.5 meV with an on-target spot size of approximately 200 μm horizontal (in the direction of the LJ propagation) and 80 μm vertical was implemented at the LJ in these experiments. The He discharge lamp at FHI delivered essentially unpolarized light to the LJ via a minimally polarizing (<0.1%) monochromator system. The energy resolution was limited by the intrinsic width of the emission lines, 1 meV (He I) and 2 meV (He II), and the focal spot size was approximately 300 × 300 μm² at the LJ. The light propagation axis of the He lamp spanned an angle of ~70° with respect to the photoelectron detection axis. The associated electron analyzer resolution was better than 40 meV at a pass energy of 20 eV. In the PETRA III experiments, the synchrotron beam was circularly polarized and the electron analyzer collection axis was aligned at 50° with respect to the light propagation axis (using the same analyzer geometry as in the SOLEIL experiment). The energy resolution was calculated to be 30 meV at 250 eV, 50 meV at 400 eV, 80 meV at 650 eV, and 140 meV at 950 eV with an associated focal spot size of approximately 180 μm horizontal (in the direction of the LJ propagation) and 20 μm vertical at the LJ. In the BESSY II synchrotron experiments, the light propagation axis was aligned orthogonally to the photoelectron detection axis. The U49-2_PGM-1 beamline (BESSY II) supplied linearly polarized soft X-rays with their polarization vector in the plane of the laboratory floor. The LJ and the photon beam propagated in this plane and were mutually orthogonal. The analyzer collection axis was aligned at ~55° with respect to the synchrotron beam polarization axis. The corresponding energy resolutions were 35 meV at ~125 eV, 70 meV at ~250 eV, 120 meV at ~400 eV, and 250 meV at ~868 eV (as determined via gas-phase photoemission resolution calibration measurements) with a focal spot size of approximately $100 \times 40$ μm$^2$ at the LJ.

The aqueous solutions were injected into the interaction vacuum chamber through 25-30 μm orifice diameter glass capillaries at the tip of a LJ rod assembly. The liquid flow rate was 0.5-0.8 ml/min. In the EASI experiments, the temperature was stabilized to 10°C by water-cooling the LJ rod using a recirculating chiller. In the SOL$^3$PES experiments, the solutions were cooled to 4°C within a recirculating chiller bath, prior to delivery to the vacuum chamber via insulating PEEK tubing. Upon injection into vacuum, the LJs exhibited a laminar flow region extending over 2-5 mm, after which Rayleigh-instabilities caused them to break up into droplets, which were ultimately frozen at a LN$_2$ trap further downstream. The laminar-flow region was surrounded by an evaporating water gas-sheath in all cases, with rapidly-decaying local gas pressures spanning ~10 mbar at the solution-vacuum interface and descending to the average vacuum chamber pressures with a 1/r distance dependence from the cylindrical LJs. The laminar region of the LJs were positioned and ionized in front of the HEA entrance apertures. The liquid-vacuum interface we refer to in the text, *i.e.*, the interface region where water's density rather smoothly decreases from its liquid bulk value to that of the gas in the immediate vicinity of the surface, is thought to evolve over a single-nm length scale.[76] The associated solutions were prepared by dissolving NaI or NaCl (both Sigma-Aldrich and of ≥99% purity) in highly demineralized water (conductivity ~0.2 μS/cm) and were degassed using an ultrasonic bath. Concentrations of 30-50 mM were used for all measurements performed under biased conditions. To measure liquid water spectra under field-free conditions, a conductive electrode was introduced in the electrically conductive liquid stream and electrically connected to the analyzer. In addition, at the beginning of every experimental run, the concentration of NaCl was iteratively varied in ~10 steps to minimize the observed width of the gas-phase photoelectron peaks. Such conditions are obtained when the potential difference between the liquid jet and analyzer entrance cone is zeroed over the liquid-gas-phase sample-light-source interaction region, with field-free conditions correspondingly pertaining, at least on average. In the EASI instrument, the corresponding optimal NaCl concentration was consistently found to be 2.5 mM at a flow rate of 0.8 ml/min and a liquid jet temperature of 10°C. The LJ rods were mounted into micrometer manipulators for high-precision alignment. The average pressures in the interaction chambers were maintained between 7 × 10$^{-5}$ and 1 × 10$^{-3}$ mbar using a combination of turbo-molecular pumping (~2000 or ~2700 l/s pumping speed for water vapor in the SOL$^3$PES and EASI instruments, respectively) and two (SOL$^3$PES) or three (EASI) liquid-nitrogen-filled cold traps (up to 18000 l/s pumping speed for water vapor per trap in both instruments). The light-LJ interaction point was set at a 500 – 800 μm distance from the detector entrance orifice, either a 500 μm (SOL$^3$PES) or 800 μm (EASI) circular differential pumping aperture. In all experiments, the LJ propagation and photoelectron detection axes were orthogonal to each other. For the experiments with the grounded LJ (field-free and streaming-potential-free measurements) all surfaces in the vicinity (at least up to 4 cm away) of the LJ-light interaction point were carefully cleaned and then coated with graphite to equalize the work function of all surfaces and prevent stray potentials: This includes the LJ rod, detector cone including the skimmer, and exit capillary of the plasma-discharge lamp. The glass capillary was not coated. We made sure that all new glass capillaries were run with water for at least a day, to passivate the inner surfaces.[28]

In both the EASI and SOL$^3$PES experiments, solutions were guided though PEEK tubing all the way to the glass capillary, *i.e.*, the liquid did not come in electrical contact with the LJ rod. In the EASI experiments, the liquid flowed through a metallic grounding insert in-between the PEEK tubing prior to injection into the vacuum chamber, *i.e.*, before entering the LJ rod assembly. In the SOL$^3$PES experiments, an electrical contact to the liquid was provided by an electrically insulated platinum disc inside the jet rod just before the glass capillary. This disc was connected via an insulated wire to an external electrical feedthrough. Both methods facilitated either the electrical grounding of the liquid to the same potential as the electron analyzer via a bridge cable or the deliberate application of a bias voltage to the liquid with respect to the analyzer. We emphasize that this biased the liquid solutions directly, and no external electrodes were used. Identical results were obtained with the two LJ rods. The bias voltages were applied using highly stable Rohde & Schwarz HMP4030 voltage sources. A sketch illustrating the LJ-PES experiment for a grounded and negatively biased water jet is presented in Figure 2A and 2B (neat water) / 2C (aqueous solution), respectively.

For the Fermi-level measurements, we utilized two metallic reference samples. Firstly, a gold wire in good electrical contact and in close proximity to the LJ (expelled by the aforementioned glass capillary nozzle) was implemented. Alternatively, a grounded platinum-iridium (PtIr) disc was used instead of the glass LJ nozzle to expel the liquid through a metallic pinhole. The PtIr disc was thus in direct electrical contact with the liquid expelled as a LJ, similar to the original LJ-PES setup utilized in Ref [4]. In the SOL$^3$PES experiments, both the liquid nozzle and the gold wire were mounted together on the same manipulator assembly and were moved in unison. The metal spectrum was measured with the LJ running after slightly relocating the whole assembly to bring the gold wire, instead of the LJ, into the synchrotron and detector foci. The EASI setup instead featured a retractable gold wire on a different port. A schematic of the PES measurement from a LJ in electric contact with a grounded gold target is presented in Figure 2D. The PtIr disc was exposed to ionizing radiation through a cutout in the disc mount towards of the detector orifice; the disc was brought into the light source focus by slightly moving the rod assembly. All methods yielded the same energetic position of the Fermi level with better than 0.03 eV precision, and no changes in the Fermi-level position were detected when running different solutions.

## Results and Discussion

**The Accurate Lowest VIE of Liquid Water, $VIE_{vac,1b1}$**

We first present results obtained with the measurement schemes introduced in Figs. 1B and 2B, *i.e.*, energy referencing Method 2 introduced above. Figure 3A shows an exemplary liquid water jet full PES spectrum (in red), ranging from $E_{cut(s)}$ to the eKE maximum, recorded with a 40.814 eV (He II α) photon energy and an applied bias voltage of -20 V. eKEs are presented as recorded by the spectrometer and under the influence of the applied bias on the top abscissa, *i.e.,* the quantity measured in the experiment. On the bottom abscissa, we plot the eKE scale with 20 eV subtracted to compensate for the applied sample bias. $E_{cut(s)}$ is found at slightly smaller energies than zero eKE when the -20-eV compensation is applied. In general, the bias voltage is slightly reduced (here about ~2%) by internal resistances between the voltage source and the liquid surface inside the vacuum chamber (for example, see Figure SI-3). However, the exact cutoff position can vary widely, as the precise KE scale depends on the particular experimental conditions, including the aforementioned residual resistance, LJ flowrate, electrolyte concentration, ionizing photon flux etc. Importantly, the absolute energetic position of $E_{cut(s)}$ or any valence features in the spectrum is of no concern for our method; we specifically aim to determine energetic separations here ($\Delta E_w$ in Figure 1B), which are not affected by the effectively applied bias voltage or any other extrinsic potential. The bias must, however, be large enough to separate $E_{cut(s)}$ from $E_{cut(A)}$ (where the former may otherwise be obscured by the latter, as illustrated in Figure 1B), and be stable on the energetic scale of the eKE measurement precision and the timescale of the experiment. Whether the measured LET curve accurately reflects the true shape and intensity of the nascent electron distribution emitted by the liquid sample with respect to the characteristic valence water PES signal intensities (commonly attributed to $1b_1$, $3a_1$, $1b_2$, and $2a_1$ orbital ionization and shown in blue in the ×20 enlarged region of the spectrum), cannot be answered here. Such a determination requires careful and technically demanding calibration of the HEA transmission under the adopted conditions.[N7]

Under the -20-V bias conditions employed here in order to utilize Method 2 (see Figure 1B), most of the gas-phase water contributions are spread out over an energy range which lies below the LET of liquid water. The remaining small tail residing below the LET – accounting for less than 0.5% of the signal, depending on the bias setting and size of the ionizing light spot – has been subtracted from the data shown in Figure 3. (Note that the small signal tail below the sample cutoff feature will generally also have a secondary electron contribution created within the detection system, although modern HEAs adopt measures to minimize such parasitic signals as much as possible.) For reference, the inset of Figure 3A shows the 20-32 eV region of the valence spectrum for the grounded water jet (in green). The unbiased spectrum exhibits simultaneous gas- and liquid-phase contributions, as commonly reported in the LJ literature[4, 6] and somewhat enhanced here due to the relatively large focal spot size of the utilized He discharge lamp.

It is of interest to discuss the un-biased spectrum (inset Figure 3A) in detail. It exhibits sharp vibrationally resolved gas-phase peaks, which are generally not observed in LJ-PES experiments. Sharp spectra of gaseous molecules are readily obtained with our setups if measurements are made without the LJ installed (see, *e.g.*, Figure SI-2 B, where the gaseous $1b_1$ HOMO ionization peak was measured by flowing gaseous water into the vacuum chamber). In that case we are not concerned with any disturbing electric fields. However, in the presence of the LJ, and with associated intrinsic and extrinsic potentials and a potential gradient acting between the LJ and the analyzer, photoelectrons from the gaseous species are accelerated differently depending on their spatial point of origin, and thus the gas-phase spectrum is inevitably broadened. In other words, a sharp gas-phase spectrum measured from water molecules evaporating from the LJ is a good indicator of a vanishing electric field in the experiments that use the relatively large focal spot of our He lamp (300 μm beam diameter). Such a field-free condition is a very useful sensor that will be exploited in the present work. A point of caution, however, is that the 'sharpness' or broadening of gaseous PE features in the presence of extrinsic fields distinctively depends on experimental parameters like the spot size of the light source or experimental resolution, and is not a universal indicator of field-free conditions.[N8] The liquid spectrum measured with the -20 V bias applied is also shown in the inset of Figure 3A (red dots), negatively shifted by the bias potential for comparison. Under these experimental conditions, an essentially pure liquid water spectrum is obtained with the gas-phase contribution shifted out of the detected energy range, as explained earlier in the manuscript in the context of Figure 1. Note that due to (experimental-geometry-dependent) differences in the relative intensities of the gas *versus* liquid phase valence ionization features, the energetic positions of the liquid-phase peaks can be easily misidentified in the absence of the applied bias. Different apparent liquid peak heights in the biased and un-biased cases reflect the fact that only the $1b_1$ gas- and liquid-phase ionization signal contributions are well separated spectrally, while for all other valence ionization channels, the two contributions overlap.[4]

We next discuss the accurate determination of $E_{cut(s)}$ and the position of liquid water's lowest VIE. For the former we analyze the spectral cutoff region and the LET, presented in Figure 3B. As in Figure 3A, the measured curve is shown in red. The violet line is the tangent extracted at the low KE inflection point, which is determined from the first derivative of the LET spectrum. The tangent intersection with the x-axis determines $E_{cut(s)}$, a standard procedure in solid-state PE spectroscopy (for example, see Refs [77-82]) that is correspondingly adopted here. The associated protocol, as well as alternative approaches to defining $E_{cut}$, are described in SI Section 5 and illustrated in Figure SI-4. For comparison, we apply two additional fit functions to the data shown in Figure 3B, the Exponentially Modified Gaussian (EMG, blue curve) distribution as originally used by Faubel and coworkers to model the liquid-phase LET curves,[36] and the distribution applied by Bouchard and Carette (green curve) as originally introduced for the description of the LET in semiconductors.[83] Both of these distributions were previously adopted in the analysis of PES spectra from a stationary droplet of saturated NaCl and NaI solutions.[68] However, neither of the two functions yield appropriate fits to the narrower experimental LET curves measured in the present work with a LJ sample, unlike in Ref. [68], supporting the associated authors' conclusion that their LET is affected by considerable surfactant impurities. Such problems are clearly avoided with a flowing and replenishing LJ, where an intrinsically sharp cutoff spectral region can be accurately measured from a liquid water sample, a similar observation to that reported in Ref. [31]. We note that the cutoff position extracted through fitting one of the aforementioned functions, or an alternative simple linear fit, often depends on the user-selected fit range, whereas a derivative-based method (like the conventional tangent approach favored here) is purely determined by the data, with no free parameters. Using the tangent method, the directly measured $E_{cut}$ value in our example is determined to be $19.64 \pm 0.02$ eV (notably including the bias-induced eKE offset; compare to the top axis in Figure 3A). Again, the fact that this value is smaller than the bias applied at the power supply ($-20.000 \pm 0.015$ V) is primarily assigned to a residual electrical resistance within the LJ, which has no relevance for our method, as outlined above and further discussed below.

In order to determine the position of liquid water's lowest ionization energy, $VIE_{vac,1b1(l)}$ (pertaining to the $1b_1$ peak maximum), we fit the valence PES spectrum in accordance with the existing literature, with two Gaussians for the well-established split second ionization threshold feature, the $3a_1$ upper and lower peaks, and a single Gaussian for the other ionization features, the $1b_1$ and $1b_2$ peaks.[4, 27] Common heights and widths of the split second VIE ($3a_1$ orbital components) were implemented for spectra recorded with sufficiently high photon energy and bias applied, i.e., in spectra where those peaks were found to be undistorted (such as that shown in Figure 3A). No other constraints were imposed on the fits. For spectra with distorted peaks and elevated inelastic-scattering background, i.e., spectra recorded with photon energies less than ~20 eV, this fit procedure was not applicable (see the next paragraph). The respective fits to the Figure 3A data, here including the lowest four water ($1b_1$, two-component $3a_1$, and $1b_2$ frontier orbital) ionization contributions, are displayed in Figure 3C. Again, the red symbols show the measured spectrum, while the green curves are the individual Gaussian fit components, and the blue curve is the cumulative fit. The lowest VIE ($1b_1$) peak is centered at $49.12 \pm 0.01$ eV KE, on the as-measured KE scale (Figure 3A top axis). Here and elsewhere in the manuscript, the eKE peak errors were taken directly from the least-squares fitting outputs and represent one standard deviation with respect to the determined peak positions. Together with the known photon energy, $h\nu = 40.814 \pm 0.002$ eV, we find $VIE_{vac,1b1(l)} = h\nu - eKE_{1b1(l)} + E_{cut} = 40.814 \pm 0.002 - 49.12 \pm 0.01$ eV $+ 19.64 \pm 0.02$ eV $= 11.33 \pm 0.02$ eV.

Results from analogous analyses of water PES spectra measured at photon energies between ~15 eV and ~950 eV are shown in Table 1, and plotted in Figure 4 (blue circles). The respective PES spectra are shown in Figure SI-5 of the SI. With sufficiently high photon energies, an analogous energy referencing can be applied to the O 1s core-orbital ionization features. Although less accurate than the $VIE_{vac,1b1(l)}$ values for the reasons we discuss below, we extract an average $VIE_{vac,O1s(l)}$ = 538.10 ± 0.05 eV for a ~650 eV photon energy, 538.07 ± 0.07 eV for 867.29 eV, and 538.04 ± 0.08 eV for 950.06 eV, all of which are in excellent agreement with the previous report of 538.1 eV, with an implied uncertainty of ± 0.1 eV.[55] The error bars and error values respectively shown in Figure 4 and reported in Table 1, as well as elsewhere in the manuscript, are the cumulative result of all error sources (calculated using standard error propagation procedures), with errors other than those arising from the peak fits being the error of the photon energy determination, error in determining the cutoff energy, and error associated with the bias-voltage shift compensation, if applied.

We make three major observations from the overall photon-energy-dependent $VIE_{vac,1b1(l)}$ data shown in Figure 4: (i) over the large photon energy range spanning 30 – 400 eV, we extract $VIE_{vac,1b1(l)}$ values between 11.31 – 11.34 eV (associated with our minimum error $VIE_{vac}$ determinations, see Table 1), (ii) for photon energies ≤30 eV, we observe an apparent significant steady increase of $VIE_{vac,1b1(l)}$ values (accompanied by increasing error bars), and (iii) the data indicate a trend towards slightly lower $VIE_{vac,1b1(l)}$ values for photon energies ≥650 eV. We start by commenting on the larger error bars determined at high photon energies (which is one aspect of point (i)). At higher soft X-ray energies, a lower overall photon flux is often combined with a rapidly decreasing photoionization cross section, requiring increased signal integration times (increasing the risk of time-dependent changes of the acquisition conditions) or compromises in the implemented acquisition settings (resolution etc.) needed to record sufficiently high signal-to-noise ratio data. Additionally, photon energies must be determined under the implemented experimental conditions, with highly precise photon-energy calibrations required when higher photon energies are used. Such processes require utmost care and still generally result in photon-energy and peak-position determinations with higher absolute errors when compared to lower-photon-energy measurements. A detailed discussion of the challenges involved in accurate photon-energy calibration can be found in SI Section 3. Another important effect to consider is the impact of the bias voltage on the detection system. A bias of several tens of volts is in effect a disturbance of the precisely tuned electron optics of a hemispherical energy (and for that matter any alternative electron) analyzer. Indeed, investigating the change in eKEs measured with our HEA systems, we find that VIE values for measurements of large eKEs can be slightly affected by the bias, depending on the bias voltage and initial kinetic energy value of the photoelectron. Specifically, it was determined that eKE values are altered by 0.015-0.035% at a bias of -64 V, depending on experimental conditions and geometric details. Figure SI-6 showcases this effect by plotting the measured $VIE_{vac,1b1(l)}$ dependence on the applied bias for exemplary measurements of the lowest-energy VIE at a photon energy of ~123.5 eV. While this effect is barely noticeable at smaller photon energies, it can become detrimental to measurements at very high photon energies, resulting in several 100 meV deviations if not properly corrected for. We have corrected all values recorded above a 100-eV photon energy by either measuring the bias-voltage dependent peak-positions directly or by additionally measuring the spectrum of the same PE feature using the residual second harmonic of the beamline and comparing it to the independently calibrated photon energy used in the measurement. This yields a correction factor for the $VIE_{vac}$ values (see SI Section 6 for details). Finally, we note that, even without such bias-voltage induced shifts, the KE-linearity of the utilized spectrometer may be a concern when the eKEs of the measured features are far apart. In our measurements, we estimate a maximal error of ~18 meV for the 950-eV measurements. If very high energy accuracy is required, then the linearity of the spectrometer eKE scale should be energy-calibrated, *e.g.*, by measuring known gas lines over a broad range of eKEs.

The apparent increase of $VIE_{vac,1b1(l)}$ values (point (ii)) for the lower photon energies is an *artifact* caused by a change of electron scattering cross sections and ionization mechanisms when tending towards lower electron KEs. For the corresponding eKEs, below ~18 eV, the direct photoelectrons experience such severe scattering that the nascent photoelectron peak position cannot be reliably extracted.[30] However, we deliberately include these misleading values in Figure 4 to highlight to the reader that utmost care must be taken when trying to determine any meaningful energy in this regime. Solely applying an energy referencing scheme, be it Method 1 or 2, without consideration of possible energy shifts due to electron scattering, will inevitably lead to erroneous results. We note that the full fit of all valence ionization features is not possible for spectra measured below 30 eV photon energies since spectral features have been considerably distorted by scattering. Accordingly, a simpler fit extracting only the lowest-ionization-energy liquid-water peak position was instead employed in that photon-energy range.

From here on, we will restrict our discussion to the meaningful photon energies at and above ~30 eV. As shown in Figure 4, and relating to point (iii) above, the precisely measured $VIE_{vac,1b1(l)}$ value determined using Method 2 in the present work is 11.33 ± 0.03 eV, which is the mean value based on the bold entries in Table 1 (corresponding to the plateau, *i.e.*, energies higher than ~30 eV but excluding the results at 650-eV photon energies and above). The new value is in very good agreement with previous values reported by Kurahashi *et al.*[28] (green squares in Figure 4) obtained using soft X-ray photon energies and the traditional Method 1 procedure, depicted in Figure 1A. This implies that in the experiments of Kurahashi *et al.* all extrinsic surface potentials including electrokinetic charging have been accurately compensated. Indeed, as further discussed below, our own carefully implemented field-free measurements based on energy referencing Method 1 allows us to extract fully consistent $VIE_{vac,1b1(l)}$ values of 11.39 ± 0.08 eV at a 40.814 eV photon energy (see Fig 5B). Comparison of our Method 1 and Method 2 results with the Method 1 results of Kurahashi *et al.*[28] and Thürmer *et al.*[84] accordingly indicates that any photon-energy dependence of $VIE_{vac,1b1(l)}$ is rather small (related to point

(iii) above). These comparisons also suggest that any effect of an intrinsic liquid-water surface-dipole potential is negligibly small or can be adequately compensated by implementing a specific electrolyte concentration that engenders field-free conditions, at least with a cylindrical liquid-microjet source. That is, in our implemented measurement geometry, any differences between $E_v^{loc}$ in the vicinity of (nearly) neat liquid water and $E_v^\infty$ seem to be below our detection limit. Considering the maximum uncertainty with which $E_{cut}$ is defined in our high energy resolution data (see SI Section 5) and stressing that direct experimental measurements of the interfacial dipole potential, $\chi^d$, have yet to be reported, our error bars support a <50 meV value of $\chi^d$, in agreement with previous experimental inferrences.[57, 58] On a related note, assuming a negligible value of $\chi^d$, the consistency of our Method 2 and properly recorded Method 1 results reinforces the use of the tangent approach to determine $E_{cut}$ from an appropriately recorded LET spectrum. Were we to adopt the inflection point of the LET curve as the $E_{cut}$ value instead of the tangent intersection point with the x-axis, we would determine just 30-100 meV higher $VIE_{vac,1b1(l)}$ values (again see SI Section 5). Focusing on our high energy resolution results recorded between 40.813 eV and ~401 eV, these offsets are limited to 30-60 meV. Thus, adopting the alternative and non-standard inflection point $E_{cut}$ definition, this would result in average and upper limit values of $VIE_{vac,1b1(l)}$ of 11.38 ± 0.03 eV and $VIE_{vac,1b1(l)}$ of 11.41 ± 0.03 eV, respectively.

Our $VIE_{vac,1b1(l)}$ results clearly disagree with the most recently reported value from Perry *et al.*[29], 11.67 ± 0.15 eV (shown in red in Figure 4). These results were based on Method 1 but were extracted by applying a small (+0.6 V) compensating bias between the jet and time-of-flight electron analyzer, under conditions where the amount of salt was not adjusted to compensate electrokinetic charging. In contrast to the originally implemented variant of Method 1, this biasing procedure seemingly has the benefit of enabling liquid-phase PES energy referencing while lifting any constraints on the concentration or type of solute under investigation (under the proviso the solution remains sufficiently conductive). In principle, assuming sufficient care is taken to mitigate all possible perturbing potentials with the bias and to appropriately calibrate the spectrometer, this should lead to the same final result as the electrolyte tuning Method 1 scheme. However, this is obviously not the case, and the large $VIE_{vac,1b1(l)}$ value determined by Perry *et al.*[29] – approximately 0.3 eV higher than all of those previously reported – probably arises from a combination of inaccurate charge compensation, additional fields caused by the applied bias, and/or the aforementioned electron scattering issues. It is difficult to quantify the relative weight of these contributions *a posteriori*. We emphasize that an attempt to compensate fields by applying a bias voltage may lead to considerable eKE offsets if not properly accounted for during calibration of the energy axis of the employed (ToF) spectrometer, as demonstrated by Nishitani et al.[54] In fact, for an applied bias voltage of +0.6 V, the determined energy offset reported in Ref. [54] for $NaBr_{(aq)}$ yields 0.25 eV, which would push the result of Perry *et al.* down to 11.42 ± 0.15 eV, a value well agreeing with our results (see Figs. 4 and 5B) and those of Kurahashi *et al.*[28]. Note that the average $VIE_{vac,1b1(l)}$ value of 11.33 ± 0.03 eV found in the present work also notably disagrees with the 11.16 ± 0.04 eV reference value (green square) measured more than 15 years ago at intermediate 60-100 eV photon energies within the range spanned in the present study (this is the first LJ-PES reference value reported by one of the present authors).[4] A likely reason for the offset of the original 11.16 ± 0.04 eV value is again a small effect of uncompensated electrokinetic charging at a time before a precise streaming potential characterization[28, 53] was established.

We next consider photon-energy-dependent variations of the $VIE_{vac,1b1(l)}$ value in more detail. The present study is the first to apply a broad range of photoexcitation energies, connecting the UV to the soft X-ray regime. Naturally, it is intriguing to explore the possibility that $VIE_{vac,1b1(l)}$ may not be exactly the same for surface water molecules and those existing deeper into the bulk solution. The probing depth into an aqueous solution is thought to be at its smallest at around 60-150 eV KE, where the electron inelastic mean free path (IMFP) curve seemingly exhibits a shallow minimum, and rises towards higher energies.[66, 67] Correspondingly, deeper probing into the solution should be enabled at higher photon energies. This raises the barely addressed question whether $VIE_{vac,1b1(l)}$ is eKE-dependent, following the eKE-dependent IMFP in water. Indeed, the observed slight variation in our extracted values – together with the values of Kurahashi *et al.*[28] – do not exclude this possibility; the IMFP from Ref. [85] is plotted as a right-hand y-scale in Figure 4 as a guide to the eye. We note that the ~50 meV larger $VIE_{vac,1b1(l)}$ value computed at the aqueous interface with respect to the liquid bulk[13] is consistent with the interfacially-sensitive 125 eV and predominantly bulk-sensitive 650 eV and higher photon energy results reported here. Unfortunately, our current error bounds do not allow us to confirm such an offset though. Based on all available data, the corresponding error bars, and the good agreement between the blue and green data points in Figure 4 – respectively measured at the low- and high-KE side of the IMFP minimum – it is argued that the KE-dependence of $VIE_{vac,1b1(l)}$ is indeed small, specifically less than 130 meV.

**Changes of Solvent VIE & Solute VIE Values in Aqueous Solutions**

Following the exact same Method 2 protocol as described above for neat water, the measurement of $VIE_{vac,1b1}$ of an aqueous solution (denoted $VIE_{vac,1b1(sol)}$) is straightforward. Aqueous solute VIEs (denoted $VIE_{vac,solute}$) are also readily determined, without assumptions. Such measurements are founded on the schemes introduced in Figs. 1C and 2C.

Figure 5A compares the neat water valence PE spectrum with that of $NaI_{(aq)}$ at 2 M concentration and tetrabutylammonium iodide (TBAI), a surfactant, at concentrations of 12.5 mM and 25 mM. These TBAI concentrations yielded approximately one half and one full monolayer (ML) of $TBA^+$ coverage at the solution surface, respectively.[86] We note that the 25 mM TBAI concentration yields approximately the same iodide surface concentrations as obtained in 2 M NaI solutions.[86] The photoelectron spectra, including the LET and leading valence features, were again measured with a 40.814 eV photon energy, the applied bias voltage was -30 V. The spectra are aligned such that the cutoff position (determined by the tangent method) falls at eKE = 0 eV. The bottom axis thus displays the eKEs following their traversal of the solution's surface. We emphasize once more that the measured energy position of the leading photoelectron peaks or $E_{cut}$ alone has no meaning, since solutes may induce several additional potentials which can arbitrarily shift all eKEs associated with different PE features. We also re-emphasize that the effectively applied bias value is not and does not need to be precisely known. The only relevant property in Method 2 is the energetic distance (and changes of this distance) between $E_{cut}$ and a peak of interest, exemplified by $\Delta E_w$ in Figs. 1B and 1C. The inset in Figure 5 shows LET features of the same data as shown in Figure 5A but instead with the water $1b1_{(l)}$ peaks aligned; note that this corresponds to the previously adopted and unsatisfactory practice of energy-referencing aqueous solution LJ-PES data to predetermined neat water $1b1_{(l)}$ VIE values. Changes in the overall spectral energy widths now appear as a shift of the cutoff position; both Figure 5A and the inset presentations are equivalent. Adopting the cutoff spectral positions, the $VIE_{vac,1b1(l)}$ energy scale (top axis) can now be referenced from $E_{cut}$ via the precisely known photon energy. Associated solute-induced changes in the water electronic structure are discussed first, and we later focus on the lowest solute ionization channel, *i.e.,* that attributed to the first I⁻ 5p atomic orbital which corresponds to the PE features at ~33 eV eKE.

When switching from neat water to 2 M NaI, a small and statistically insignificant (*i.e.*, within the error bars) energy shift, accompanied with a slight broadening, of the $1b_1$ peak is observed with respect to neat water; see the purple and blue curves (Figure 5A). This is a somewhat puzzling result, seemingly at odds with theoretical works on alkali-halide solutions, specifically reporting a larger surface propensity of iodide than the sodium cation, which implies formation of an interfacial dipole.[76] Interestingly, the aforementioned work by Tissot *et al.* makes a related observation. Comparing concentrated NaCl and NaI aqueous solutions, which should exhibit a very different surface potential, no differences are found in the spectra;[68] those authors discussed the possibility of surface impurities obscuring their results. We note that the 2 M NaI concentration used here may still be below the surface-enrichment regime,[87] and higher concentrations (>6 M) may in fact lead to a more pronounced shift. However, a concentration-dependent study is beyond the scope of this work. If simple alkali-halide salts do not alter the solution's charge equilibrium (and thus the position of $E_{cut}$ and the valence ionization features) at the probed interface, one must assume that inter-ionic dipoles have no net component perpendicular to the solution interface. Unfortunately, there is little data available to clarify this issue, despite multiple works attempting to quantify the interfacial density profiles of the different atomic ions in aqueous solutions.[38, 76, 88, 89] In this context, some of the authors have recently reported that concentrated electrolytes, despite changing the electronic structure of water, do not appear to lead to any significant relative energy shifts between different valence photoelectron peaks.[7] Rather, the lowest-energy ionization ($1b_1$) peak slightly broadens, with an accompanied apparent narrowing or energy-gap reduction of the split second ionization ($3a_1$) feature, the latter being the more notable spectral change. Both of these behaviors are confirmed in the present data shown in Figure 5A.

Compared to the NaI results, the TBAI aqueous solutions behave very differently, shifting water's valence electronic structure with respect to $E_v^{loc}$, as reflected in the higher measured kinetic energies (green and orange curves). This energy shift is approximately 630 meV, judged from the change of the neat water $1b_1$ peak position, in the case of a full ML of TBA$^+$ (compare the green and blue spectra). A coverage of 0.5 ML leads to a smaller shift of about 530 meV (orange spectrum). We thus find average VIE$_{vac,1b1(TBAI)}$ values of 10.80 ± 0.05 eV (0.5 ML) and 10.70 ± 0.05 eV (1.0 ML), which are both found to be considerably smaller than VIE$_{vac,1b1(l)}$. This large decrease in VIE could have various causes: (1) resulting from changes of the intrinsic (bulk) electronic structure of the solution (as shown for NaI), (2) a change of the intrinsic electronic structure and associated charge equilibrium at the solution-vacuum interface (i.e., a relative change in the positions of water's electronic bands with respect to a fixed value of $E_v^{loc}$), or (3) a change in the net aqueous surface-dipole potential and associated value of $E_v^{loc}$. A change of the bulk-water electronic structure would be hardly expected for this surface-active molecule. However, we may have to consider the possibility of changes of the aqueous electronic structure at the liquid – vacuum interface. Still, such an effect would need to be distinguished from the two other interfacial contributions, requiring establishment of a common and (ideally) ion-depth-invariant reference level for the two solutions. The Fermi level should be well-suited to this task and can be indirectly measured using the experimental procedure discussed in the context of Figure 1D. However, before discussing such a referencing procedure in detail, we consider the iodide solute signal, as measured with respect to $E_v^{loc}$, which is also visible in the spectral range displayed in Figure 5.

Iodide photoemission gives rise to the small I$^-$$_{(aq)}$ 5p doublet features (multiplied here by a factor of 100) occurring in the 32.0-34.4 eV KE region in Figure 5. Applying a 2-Gaussian fit procedure, we determine the respective peak positions at eKEs of ~33.6 eV (I$^-$ 5p$_{3/2}$) and ~32.7 eV (I$^-$ 5p$_{1/2}$) in the case of a 1 ML TBAI$_{(aq)}$ solution. Slightly lower eKEs of ~33.4 eV and ~32.5 eV are determined for a 0.5 ML TBAI$_{(aq)}$ solution. This corresponds to a VIE$_{I5p3/2}$ = 7.20 ± 0.1 eV / VIE$_{I5p1/2}$ = 8.11 ± 0.1 eV for the 1 ML and VIE$_{I5p3/2}$ = 7.38 ± 0.1 eV / VIE$_{I5p1/2}$ = 8.30 ± 0.1 eV for the 0.5 ML cases, respectively. In contrast, for a 2 M NaI$_{(aq)}$ solution we find an eKE of ~32.7 eV (I$^-$ 5p$_{3/2}$) and ~31.8 eV (I$^-$ 5p$_{1/2}$), corresponding to VIE$_{I5p3/2}$ = 8.08 ± 0.1 eV / VIE$_{I5p1/2}$ = 8.90 ± 0.1 eV; the latter is in excellent agreement with our earlier work[2]. An important finding from Figure 5A is, therefore, that the iodide 5p ionization energy is considerably larger in the NaI aqueous solution as compared to the TBAI solution. We note that the observed effect would have been much smaller if we had used Method 1, where only the VIE$_{vac,1b1(l)}$ - VIE$_{I5p}$ energy distance would be accessed but not the change of VIE$_{vac,1b1}$. While this energy separation is indeed different by about ~0.1 eV between 0.5 ML and 1.0 ML TBAI and about ~0.25 eV between 2 M NaI and 1 ML TBAI (as could have been observed via Method 1), the true change of VIE$_{I5p}$ as determined with Method 2 would remain inaccessible. Notably, a pervious study[90] used gas-phase water features as an energy reference for 0.04 m TBAI$_{(aq)}$ solution PES, and thus circumvented the liquid $1b_1$ VIE altogether, arriving at rather accurate VIE$_{I5p3/2}$ = 7.6 eV and VIE$_{I5p1/2}$ = 8.4 eV values, albeit with a potentially huge margin of error due to unknown and uncompensated extrinsic potentials. Specifically, for NaI$_{(aq)}$, the energetic separation of water's lowest ionization energy 1b$_1$ peak to the I$^-$ 5p$_{3/2}$ peak is 3.36 ± 0.05 eV, while for the 5p$_{1/2}$ peak it is 2.41 ± 0.05 eV, in excellent agreement with previous reports.[28, 68, 87] Figure 5 also shows that VIE$_{vac,1b1(TBAI)}$ is slightly smaller in the case of 1.0 ML TBAI coverage in comparison to 0.5 ML coverage. However, the associated energy difference is smaller than the respective changes in the VIE$_{I5p}$ energies. For 1.0 ML TBAI solutions, we see a ~0.25 eV increase in the water 1b$_1$ to I$^-$ peak separations in comparison to the 2 M NaI case. This corresponds to 3.60 ± 0.05 eV and 2.65 ± 0.05 eV separations of the 5p$_{1/2}$ and 5p$_{1/2}$ peaks to the water 1b$_1$ peak, respectively. This aspect will be considered further below.

We close this sub-section by re-connecting the results reported here to the applicability of Method 1. Figure 5B presents additional PES spectra from neat water and 1.0 ML TBAI, now measured for grounded solutions, *i.e.*, applying Method 1. We observe the very same positions of VIE$_{vac,1b1(l)}$ as in the upper trace, obtained with Method 2. The reason for this (perhaps surprising) quantitative agreement is that in this particular Method-1 measurement all external fields were successfully compensated. This is true for both neat water and the TBAI solution spectra as judged by the sharp water-gas-phase features; we re-emphasize that the extrinsic fields between the sample and analyzer can only be meaningfully assessed when a sufficiently large gas volume around the LJ is probed. Establishing the necessary field-free conditions to achieve such measurements is however experimentally difficult and time-consuming. More importantly, these conditions are impossible to achieve for most aqueous solutions outside a very limiting concentration range, and only Method 2 will undoubtedly provide the correct ionization energetics. In the next section, we extend our newly applied energy referencing methodology a step further, establishing a common Fermi level for neat water and a metallic reference sample that allows determination of the VIE of liquid water with respect to E$_F$, VIE$_{EF,1b1(l)}$. Furthermore, in combination with the Method 2 results, it provides access to the liquid water eΦ value, eΦ$_{water}$.

**Fermi-Referenced VIEs & Work Functions of Liquid Water & Aqueous Solutions**

As argued when describing Figs. 1D and 2D, we can formally introduce Fermi-level referencing when liquid water or an aqueous solution is in electrical equilibrium with a metallic reference sample, an approach we term Method 3. As explained in the introduction, we can measure the valence spectrum from a solution and E$_F$ from a metal in sequential experiments. But exactly what information does this provide? With the two systems in electrical contact, E$_F$ and the bulk chemical potential of the solution and the metal are aligned. However, in the PES experiment, one measures photoelectrons from the solution or metal after they have traversed the sample-vacuum interface and different corresponding surface dipole potentials. Ideally, one would measure the Fermi edge of the metal through a thin sheet of the flowing solution, such that electrons emitted from the metal and the (bulk) solution would experience the same (intrinsic and extrinsic) solution-vacuum surface potential and E$_v^{loc}$. However, as of yet, this remains experimentally unfeasible.[N9] Despite this, it can still be argued that a Fermi-level alignment can be achieved between the LJ and metallic reference if *streaming-potential-free* conditions are engineered, *i.e.,* under the experimental conditions depicted in Figure 1D. We define these conditions as those that preserve the intrinsic liquid solution $\Delta e\Phi$ value with respect to the analyzer, while mitigating the remaining extrinsic potentials. It is important to differentiate these conditions from the field-free alternative discussed in the context of Method 1, where the sum of *all* potentials between the sample and analyzer are compensated to zero. This point is particularly noteworthy as the establishment of field-free conditions has previously been symbolized as '$\Phi_{str} = 0$'.[28] In general, the optimal solution concentrations for field-free and streaming-potential-free conditions differ, offset by the magnitude of $\Delta e\Phi$ in the experiment. Only if $\Delta e\Phi$ happens to be zero (for a particular experiment) will these two conditions be simultaneously achieved (at a fixed LJ nozzle morphology, jet flow rate, and solution temperature).

In the following, we briefly discuss how streaming-potential-free conditions may be established by considering the streaming current of the aqueous sample, $I_{str}$, which is the source of the streaming potential, $\Phi_{str}$, and a less ambiguous quantity. $I_{str}$ has been measured independently from $\Phi_{str}$, where it was shown that the aqueous streaming current is minimized at roughly 50 mM alkali halide salt concentrations, with a LJ flow rate of 0.5 ml/min, and with similar LJ nozzle orifices as implemented here.[28, 53] Accordingly, a 50 mM NaI salt concentration provides a basis for our Fermi-referencing measurements.[N10] Associated nominally streaming-potential-free liquid water PES results recorded with a photon energy of 40.814 eV are shown in Figure 6 (blue curve). At higher eKEs, we show the related Fermi-edge spectrum of the metal reference sample (black curve) sequentially recorded under the same conditions, as sketched in Figure 2D. Here, the liquid water jet was in operation in close proximity to a gold wire or was directly injected from a conductive PtIr disc during these measurements. An associated fit to a Fermi-Dirac distribution is also shown (lilac curve), from which we obtain the Fermi-edge position at $eKE_{EF} = 36.296 \pm 0.005$ eV. This position defines the zero of the $VIE_{EF}$ energy and chemical potential scale (lower axis at the top of the figure), to which all liquid-water features can now be referenced. The difference between $VIE_{vac,1b1(l)}$ and $E_F$, as determined from our fits, corresponds to $VIE_{EF,1b1} = 6.60 \pm 0.08$ eV. Analogous measurements were performed using $125.02 \pm 0.03$ eV, and $649.946 \pm 0.005$ eV photon energies and yielded the same results.

To examine whether streaming-potential-free conditions were established when recording the liquid water data shown in Figure 6, and the associated validity of the measured $VIE_{EF,1b1}$ value, a series of aqueous-phase PES spectra in electrical contact with a grounded metallic reference sample were recorded as a function of salt concentration (NaI and NaCl were found to exhibit the same effects). This allowed us to track the shift of aqueous-phase PE features with respect to $E_F$. The resulting spectra are plotted in Figure SI-7, with the energetic position of the Fermi level found to be fixed within 0.03 eV, regardless of the type of aqueous solution present. That is, the metallic spectrum appears to be unaffected even by relatively high extrinsic potentials at the LJ (in some cases exceeding 1 eV). One may speculate that such potentials are effectively screened and thus terminated at the metal, nullifying any field gradients in the region between the metal and the detector. However, in contrast in the liquid water case, the lowest VIE $1b_1$ feature shifts dramatically under the influence of the varying salt concentration and streaming potential, displaying the expected behavior and exhibiting a minimum $VIE_{EF,1b1}$ value around 50-100 mM concentrations, *i.e.*, covering the concentration implemented to produce the blue curve in Figure 6 and where $I_{str}$ (and in turn $\Phi_{str}$) is expected to vanish.[N11] This implies that streaming-potential-free conditions have indeed been achieved in producing the liquid water data shown in Figure 6.

Recalling our aforementioned determination of the $VIE_{vac}$ energy scale of liquid water using Method 2 (see the upper axis above Figure 6), we are now set to relate the vacuum and $E_F$ energy scales to each other. Since $e\Phi$ is equivalent to the difference between these two energy scales, *i.e.*, between the ionization energies of any of liquid water's ionization features measured with respect to $E_v^{loc}$ and $E_F$, we can accordingly determine $e\Phi_{water}$. For example, $VIE_{vac,1b1(l)} - VIE_{EF,1b1(l)} = 11.33 \pm 0.03 - 6.60 \pm 0.09$ eV and yields $e\Phi_{water} = 4.73 \pm 0.09$ eV. By extension, one can further argue in the case of neat water that if the surface dipole / outer potential is zero, near-zero, or averages to zero in our experiments, then $e\Phi_{water} \approx \bar{\mu}$, *i.e.*, the determined work function is equivalent to water's electrochemical potential $\bar{\mu}$, which is a generally un-measurable quantity. We again stress that without establishing streaming-potential-free conditions, arbitrary $VIE_{EF,1b1(l)}$ and thus $e\Phi_{water}$ values would be recorded, depending on the strength and sign of any extrinsic potentials; as demonstrated by Figure SI-7. Generally, this will remain a problem whenever the metallic reference spectrum is measured separately from the solution spectrum (*i.e.*, unless the Fermi edge signature and liquid features of interest are recorded in the same spectrum, following ejection *through* the liquid surface). This issue is unfortunately somewhat obscured when the metallic reference sample is used to initially establish an alternative (but nonetheless flawed) reference, such as the analyzer work function, as proposed, *e.g.*, in Refs. [31, 69].

As a further cross-check of our $VIE_{EF,1b1}$ and $e\Phi_{water}$ results, and that streaming-potential-free conditions are indeed achieved, we extract and utilize our analyzer work function, $e\Phi_A$. To achieve this, we measured PES spectra of the metallic reference sample, either directly recording the Fermi level ($e\Phi_A = h\nu - KE_{EF}$) or some other well-calibrated metal energy level such as the gold 4f level ($e\Phi_A = h\nu - BE_{4f} - KE_{4f}$). The extracted $e\Phi_A$ is an arbitrary value in itself, and only equals the analyzer work function if the measured kinetic energy, $eKE_{meas}$, of the detection system has been precisely calibrated using known (gas-phase) reference photon and ionization energies. We briefly describe the procedure to achieve such a calibration and compare the $e\Phi_A$ result to the field-free condition, specifically assuming this corresponds to $\Delta e\Phi = -e\cdot\Phi_{str}$. Using the kinetic energy position of the equilibrated water gas-phase $1b_1$ peak (compare to Figure SI-2) and the associated reference VIE value of $12.621 \pm 0.008$ eV[3], we find that the kinetic energy scale of the detector needs to be corrected by $+0.224 \pm 0.008$ eV (note that this value depends on the pass energy setting and detector mode). This yields a corrected Fermi-edge position of $eKE_{EF}$ = 36.520 ± 0.009 eV from which we determine $e\Phi_A$ = 4.293 ± 0.009 eV, a value approximately 0.43 eV smaller than $e\Phi_{water}$. It is intriguing to then compare this value to the shift in the liquid water $1b_1$ position when going from our streaming-potential-free (50 mM) to field-free conditions (2.5 mM), *i.e.*, where $\Delta e\Phi$ = -$e \cdot \Phi_{str}$. There we observe that the $1b_1$ peak shifts to lower eKEs (compare to Figure SI-7) and that the overall shift between these two concentrations matches the expected 0.43 eV. This nicely demonstrates the shift from $\Phi_{str}$ = 0 V conditions to $\Delta e\Phi$ = -$e \cdot \Phi_{str}$ conditions, that the liquid water $1b_1$ peak follows the change in potentials one-to-one, and that streaming-potential-free conditions were indeed achieved with 50 mM NaI concentrations (under our implemented conditions). Correspondingly, our values of $VIE_{EF,1b1}$ and $e\Phi_{water}$ are also confirmed.

Our established experimental value of $e\Phi_{water}$ is found to be somewhat larger than that reported by Olivieri *et al.*[31], 4.65 ± 0.09 eV, who also attempted to determine $e\Phi_{water}$ using LJ-PES. This work extracted the value of $e\Phi_{water}$ from the 'midpoint' of the rise of the LET curve (referred to as the SEED in the Ref. [31], see Note N3) as opposed to the tangent method commonly adopted for solid state samples and in extracting the results reported here. In our data, this $E_{cut}$ determination method has been shown to result in VIE increases of several 10 meV up to ~150 meV (depending on the energy resolution of the experiment and the associated shape of the LET). This would be directly transferred to an increase of our value of $e\Phi_{water}$, bringing our determination of this value further away from that reported by Olivieri *et al*. With a comparison of these and our own $e\Phi_{water}$ value determinations in mind, we highlight a number of methodological inconsistencies and inaccuracies in the Olivieri *et al.*[31] study in SI Section 7.

Turning now to an attempted determination of $E_F$ and $e\Phi$ from an aqueous solution, we recall that our auxiliary Fermi-referencing procedure (Method 3) is not applicable to, *e.g.*, the 2 M NaI solution considered in Figure 5, as the streaming current is thought to be non-zero (see SI Figure 7). Although a precise value cannot be determined in this work due to the coupling of higher salt concentrations to $\Phi_{str}$, we can compare $VIE_{EF,1b1(l)}$ (*i.e.*, from water) with the respective value from Tissot *et al.*[68] for saturated alkali-halide solutions deposited on a gold substrate. There a 0.4-eV smaller $VIE_{EF,1b1(sol)}$ of 6.2 eV was reported. However, we note that for higher concentrations of 2 M NaI (Figure 6) and 4 M NaI (Figure SI-7) the $1b_1$ peak notably shifts to higher $VIE_{EF}$ (lower eKEs) values, *i.e.*, even further away from the reported 6.2 eV $VIE_{EF,1b1(sol)}$ value. The shift observed in our high-concentration measurements is likely caused by a non-zero $\Phi_{str}$, and one can only speculate about the true $VIE_{EF,1b1(sol)}$ value in the absence of $\Phi_{str}$. However, a value of 6.2 eV is deemed unlikely. We may speculate that, to some extent, this 6.2 eV determination reflects additional extrinsic surface potentials present at the interface of the concentrated solution in the Tissot *et al*. study[68]. This is consistent with an observed ~0.6 eV energy shift of the O 1s gas peak towards lower eKE (higher VIE) when retracting the sample,[68] caused either by radiation-induced sample changes or accumulation of surface impurities of the non-replenishing solid sample.

We now return to our TBAI aqueous solution measurements, where we observed large changes in VIE$_{vac}$. At a bulk concentration of 25 mM, the solution conductivity is sufficient to effectively apply a bias voltage of -30 V, and we can assume alignment of E$_F$ throughout the solution, similar to the 25-50 mM NaCl or NaI aqueous solution cases discussed above. Consequently, we can determine E$_F$ following the same steps as for neat water. For that we reproduce the TBAI aqueous solution spectrum from Figure 5B in Figure 6 (green curve), and compare it to the Fermi edge spectrum from the metallic sample (black curve), fully analogous to the water experiment. As discussed above, even when measured in the presence of the running TBAI-solution jet, electrically connected to the metallic sample, the same E$_F$ reference value is observed as for neat water. Neither I$_{str}$ nor Φ$_{str}$ measurements have been reported for this solute to our knowledge, and are beyond the scope of this work. However, as we argue in the following, Φ$_{str}$ may in fact be immeasurably small or even zero in this particular case. Before explaining this further, we discuss the principal results from the green curves in Figure 6. Initially assuming Φ$_{str}$ ≈ 0 V, we determine that water's 1b$_1$ PE peak shifts to 0.15 ± 0.11 eV lower Fermi-referenced VIE values in the TBAI$_{(aq)}$ solution in comparison to (nearly) neat water, with VIE$_{EF,1b1(TBAI)}$ = 6.45 ± 0.08 eV. Using the results from Method 2 we can now, analogously to the water case, determine the solution's work function: eΦ$_{TBAI}$ = VIE$_{vac,1b1(TBAI)}$ − VIE$_{EF,1b1(TBAI)}$ = 10.70 ± 0.05 − 6.45 ± 0.08 eV = 4.25 ± 0.09 eV. This corresponds to a 0.48 ± 0.13 eV reduction with respect to neat water. Considering the anionic solute components of the solution, we further extract VIE$_{EF,I5p}$ values of 3.80 ± 0.10 eV and 2.84 ± 0.10 eV for the 5p$_{3/2}$ and 5p$_{1/2}$ peaks of the I$^-$ solute feature, respectively.

We have seen that field-free conditions are seemingly achieved for 25 mM TBAI solutions (implied by the sharp water gas-phase spectrum in Figs. 5B and 6), which must mean eΦ$_{TBAI(aq)}$ - eΦ$_A$ ≈ -Φ$_{str}$. Recalling that eΦ$_A$ = 4.293 ± 0.009 eV, it follows that eΦ$_{TBAI(aq)}$ = 4.25 eV ≈ eΦ$_A$, *i.e.*, the E$_v^{loc}$ levels at the sample and analyzer are aligned, implying Φ$_{str}$ ≈ 0 V. We have also observed that Φ$_{str}$ ≠ 0 V for the 2 M NaI$_{(aq)}$ solutions, as can be seen in Figure 6 from the offset of the spectrum towards slightly higher apparent VIE$_{EF,1b1}$ values. However, Method 3 does not reveal whether eΦ$_{TBAI(aq)}$ or Φ$_{str}$ is compensating the extrinsic potential, implying that the observed shift in VIE$_{EF,1b1}$ may come from an active Φ$_{str}$ and that the field-free condition achieved here is just a coincidence. The origin of the observed energy shift (change of VIE$_{vac,1b1(l)}$) in Figs. 5 and 6 thus remains unresolved, and is unanswerable with the currently available experimental tools. However, we briefly discuss how a real change in VIE$_{EF,1b1}$, *i.e.*, under the premise that Φ$_{str}$ = 0 V, would be realized below.

Dissolution of a salt in water produces hydrated anions and cations, which can be viewed as ionized dopants freely moving in the aqueous solution. At the interface to vacuum this would give rise to the band bending (BB) phenomenon commonly encountered in the semiconductor literature, and illustrated in Figure SI-1B. In the present case, BB is argued to be induced in response to TBAI accumulation, which changes the charge distribution at the liquid-vacuum interfacial layer. Briefly, BB occurs if there is a local imbalance of charge near the surface which leads to the build-up of a local field.[26, 35, 91, 92] Arguably, we observe an upward BB, *i.e.*, in the direction of lower VIEs, which is caused by a depletion of (the solvent's) electron density near the surface. The hydrophobic TBA$^+$ molecules which reside near the solution's surface are thought to draw I$^-$ ions into this surface region.[86] It can then be argued that the solvation of I$^-$ reduces water's local electronic density, leading to the observed effect. Notably, the Fermi level remains fixed (the Fermi level is pinned) within the solution at its bulk value, and aligned with the metal reference and analyzer, as shown in Figure SI-1B. Some intriguing observations support this interpretation. The 1b$_1$ HOMO peak slightly broadens when moving from 50 mM NaI$_{(aq)}$ to 25 mM TBAI$_{(aq)}$ solutions, which may indicate that an interfacial region with a solution-depth-dependent potential energy gradient of the 1b$_1$ band is probed, implying different effective 1b$_1$ energies within this so-called space-charge layer. Also, the I$^-$ 5p peaks are shifted the farthest, which would be plausible given that most of the iodide resides directly at the surface, where the most disturbance of the bulk equilibrium occurs. One might correspondingly ask whether the neat water surface is already subject to BB, keeping in mind that intrinsic surface BB caused by the presence of surface defect states is a common phenomenon for semiconductors.[92] While we cannot rule out this possibility completely, it is important to note that the water surface is very different from an abruptly terminated crystal lattice, and the dynamic nature of liquid water is likely to compensate for any charge imbalance (unless such charge accumulation is forced as in the case of surface-active species such as TBAI). Thorough exploration and characterization of such effects using photon-energy- and thus solution-depth-dependent Fermi-referenced LJ-PES measurements is an associated interesting future line of research directly enabled by the work reported here.

Until now we have adopted surface-science concepts to interrogate and interpret aqueous-phase PES data, providing a useful methodological advancement to access an explicit descriptor of solution interfacial electronic properties, namely the work function via joint determinations of VIEs with respect to $E_v^{loc}$ and $E_F$. In the following, we briefly discuss the impact of this accomplishment in the wider context of interfacial chemistry and electrochemical processes, in particular at the metal-electrode – electrolyte system. This very ensemble of a LJ electrically connected with a metal sample (again, see Figure 2D) represents a single electrode immersed into an electrolyte. As we have explained above, connection of a metal to a sufficiently conductive liquid water or aqueous solution sample (both classifiable as semiconductors) yields a common Fermi level. In the case of an electrolyte containing both forms of a redox couple (representing vacant and populated energy levels within the band gap, separated by $E_F$[15]), the redox level, $E_{redox}$, can be equated to $E_F$ in the solution and aligned with $E_F$ of the metal.[26] This implies that $E_F$ of the solution shifts with charge flow across the interface until $E_F = E_{redox}$, where the two energy scales for the aqueous solution and the potential scale for the electrode are connected through the theoretical value of the Fermi level of the standard hydrogen electrode. This route has been explored in a very recent LJ-PES study,[69] determining $E_F$ via the aqueous-phase ferricyanide/ferrocyanide redox couple (in a Zobell[70] solution), and reporting values of VIE$_{EF,1b1(l)}$ = 6.94 eV and $e\Phi$ = 4.60 eV, both of which differ from our results for neat liquid water. Furthermore, a much larger $VIE_{vac,1b1(l)}$ value of 11.55 eV was reported for the Zobell solution. We highlight a number of potential issues with the methodology adopted in Ref. [69] in SI Section 7, which we believe may be responsible for the discrepancies between our and their results. We also note that most of these problems could be circumvented by rigorously applying Method 2, as presented in this work.

In a more general context, and not requiring introduction of redox couples, it will be possible to use a known electrode potentials and measured Fermi levels to locate the band edges of liquid water and select aqueous solutions on the chemical potential scale.[26] This is not only of uttermost importance for advancing our understanding of chemical reactions at electrode-electrolyte systems but it also enables future routes to develop a common interpretation of thus far seemingly disconnected quantities specific to the molecular and condensed-matter descriptions of electronic structure. One pressing example is how the band gap of liquid water conceptually connects with the molecular-physics or orbital information accessed by LJ-PES, including an experimental determination of liquid water's electron affinity.[13]

## Conclusions

Liquid microjet photoelectron spectroscopy (LJ-PES) is an indispensable experimental tool for the characterization of electronic-structure interactions in liquid water and aqueous solutions. This includes determination of valence electron energetics, which is key to understanding chemical reactivity. And yet, the full potential of this method is just about to be exploited, entailing several important benefits, discussed in the present work. This includes the measurement of *absolute* solute and solvent energetics and the accessibility of a specific interfacial property descriptor, the work function (something that is routinely obtained in solid-state PES). Specifically, we have demonstrated the necessity of measuring the liquid-phase low-energy cutoff spectrum along with the photoelectron peak of interest. This approach has several major advantages over the formerly adopted LJ-PES energy referencing scheme and correspondingly has far-reaching implications. With the help of the cutoff energy, $E_{cut}$, absolute solute and solvent energies can be robustly, accurately, and precisely measured without assumptions, no longer requiring the long-practiced and unsuitable energy referencing to the lowest-energy $VIE_{1b1}$ of *neat* liquid water. Using the methodology introduced here, we find an average $VIE_{vac,1b1}$ of 11.33 ± 0.03 eV (with respect to $E_v^{loc}$) for neat water, and attribute several previously measured and offset values to the effects of perturbing surface charges, with various condition-dependent potential origins. Via a broad photon energy dependent study of $VIE_{vac,1b1}$, spanning the UV and a large portion of the soft X-ray range, there is a further indication of a small photon-energy dependence of $VIE_{1b1}$, although a definitive answer has to be postponed until the challenge of precisely measuring VIEs with a small error at high photon-energies can be overcome. We further demonstrated the emergent ability to measure solute-perturbed $VIE_{vac,1b1}$ values from aqueous solutions, *i.e.*, solute-induced effects on water's electronic structure. With the same experimental approach, solute energies can be accurately measured, something which is exemplified here using aqueous iodide solutions. Extending our proposed energy referencing approach to deeper-lying electronic states, we have additionally reconfirmed and more precisely defined water's O 1s core-level binding energy, extracting a value of $VIE_{vac,O1s} = 538.10 \pm 0.05$ eV at a ~650 eV photon energy.

Regarding the interfacial properties of water and aqueous solutions, we have described and applied a procedure that allows the formal determination of the Fermi level of neat water and select aqueous solutions. Our approach is based on the measurement of LJ-PES spectra under conditions where the streaming potential associated with the flowing LJ has been mitigated. It further relies on (the separate) measurement of the Fermi edge spectrum from a metal sample in good electrical contact with the electrolyte and electron analyzer. This allowed us to accurately determine $VIE_{EF,1b1} = 6.60 \pm 0.08$ eV (with respect to the $E_F$). Building on this approach and the separate accurate measurement of vacuum-level-referenced VIEs (as discussed above), interface-specific aqueous-phase work functions have been extracted, including that of liquid water. Here, $e\Phi_{water}$ was accurately determined to be $4.73 \pm 0.09$ eV. Based on the collective electronic structure information accessed both with respect to $E_v^{loc}$ and $E_F$ over the course of this study, we have carefully discussed the observed solution-specific energy shifts of the $E_{cut}$ feature and/or VIE values, which have allowed us to differentiate solution work function and solute-induced (bulk) electronic structure changes. This included quantification of a nearly 0.5 eV aqueous solution $e\Phi$ reduction upon dissolution of a known surfactant (0.025 M TBAI).

Still, our study also highlights current shortcomings in state-of-the-art liquid-phase experimental methodologies, particularly the difficulties in $E_F$-referencing arbitrary, free-flowing aqueous solutions and determining their work functions. This primarily stems from the challenges associated with mitigating solution streaming potentials, irrespective of their concentrations, surface dipole potentials, and the employed experimental conditions. In the particular case of liquid water, we have shown that the aforementioned limitations can be circumvented by measuring and zeroing streaming potentials, while taking advantage of liquid water's small $\chi^d$ value. Here, the inaccuracies of this approach have been determined to amount to less than 50 meV, notably within our $VIE_{EF,1b1}$ and $e\Phi_{water}$ error ranges. However, in the case of concentrated aqueous salt solutions, such an approach could not be adopted, specifically due to the presence of unknown streaming potentials *and* $\chi^d$ values. To overcome these limitations, an alternative and more general $E_F$-referencing method would need to be realized. An intriguing associated approach would be the detection of photoelectrons from a solid sample (specifically a metal) covered with a thin layer of flowing electrolyte, engendering metal- and solution-born electron collection via the same, generally charged liquid interface. This, however, remains a formidable challenge, particularly for PES studies aiming to resolve the microscopic (electronic) structure and chemical processes occurring at solid-solution interfaces.[65] Irrespective of the various technical hurdles ahead, the work presented here is a major enrichment of the LJ-PES technique, enabling the general, direct, and accurate measurement of absolute electron energetics within the liquid bulk and at liquid-vacuum interfaces of aqueous solutions. Concurrently, this work brings us a step closer to bridging the gap between solid-state and liquid-phase PES, and more importantly the surface science and (photo)electrochemistry research disciplines.

## Data availability

The data of relevance to this study have been deposited at the following DOI: 10.5281/zenodo.5036382.

## Conflicts of interest

There are no conflicts to declare.

## Author Contributions

S.T., B.W., and I.W. designed the experiments and, together with S.M, F.T., and with occasional assistance from C.L., performed the measurements. S.T., B.W., S.M., and I.W. analyzed the data. S.T., B.W., and I.W. wrote the manuscript and the Supplementary Information with critical feedback from all co-authors.

## Acknowledgment

S.T. acknowledges support from the JSPS KAKENHI Grant No. JP18K14178 and JP20K15229. S.M., U.H., and B.W. acknowledge support by the Deutsche Forschungsgemeinschaft (Wi 1327/5-1). F.T., G.M., and B.W. acknowledge support by the MaxWater initiative of the Max-Planck-Gesellschaft. B.W. acknowledges funding from the European Research Council (ERC) under the European Union's Horizon 2020 research and investigation programme (grant agreement No. 883759). D.M.N. and C.L. were supported by the Director, Office of Basic Energy Science, Chemical Sciences Division of the U.S. Department of Energy under Contract No. DE-AC02-05CH11231 and by the Alexander von Humboldt Foundation. We thank the Helmholtz-Zentrum Berlin für Materialien und Energie for allocation of synchrotron radiation beamtime at BESSY II and Robert Seidel for his support during the associated measurements. We acknowledge DESY (Hamburg, Germany), a member of the Helmholtz Association HGF, for the provision of experimental facilities. Parts of this research were carried out at PETRA III and we would like to thank Moritz Hoesch in particular as well as the whole beamline staff, the PETRA III chemistry laboratory and crane operators for assistance in using the P04 soft X-ray beamline. Beamtime was allocated for proposal II-20180012 (LTP). Some of the experiments were carried out with the approval of synchrotron SOLEIL (proposals number 20190130). We thank Laurent Nahon and Sebastian Hartweg in particular, as well as the technical service personnel of the SOLEIL chemistry laboratories for their assistance.

# Figures

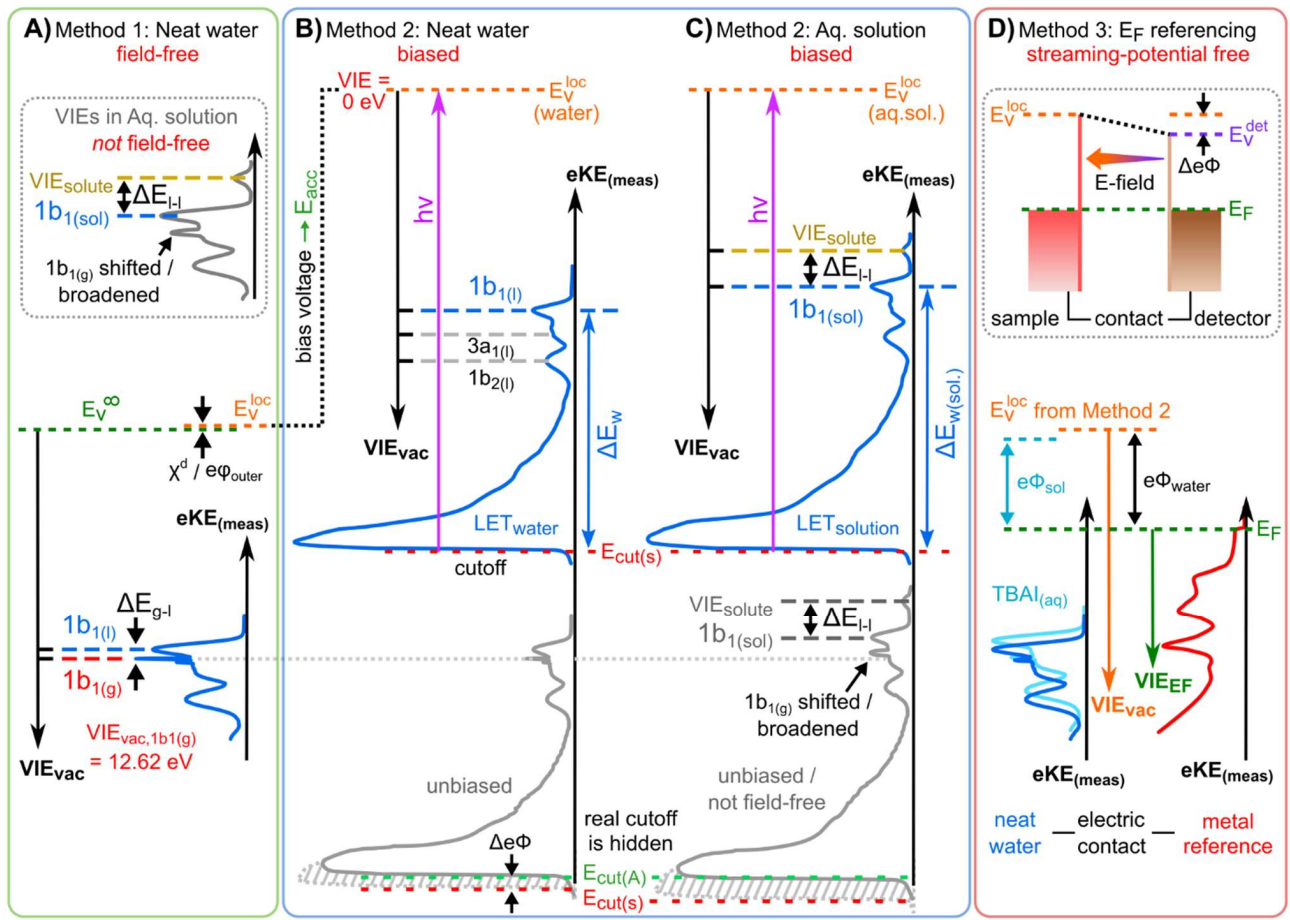

**Figure 1:** Schematic electronic energetics for each experimental method described in the main body of the text. **A)** Both gas and liquid spectral features are measured together on the eKE$_{(meas)}$ scale under field-free conditions (blue spectrum), which makes it possible to use the known gas-phase VIE values (red) as an energy reference; ionization energies, VIE$_{vac}$, are referenced to the vacuum level at infinity, E$_v^\infty$. The inset shows the commonly adopted extension to Method 1 to reference solute VIE values by determining the solute peak's energetic distance to the liquid water 1b$_1$ peak, $\Delta E_{l-l}$, and (generally inappropriately) using the VIE$_{vac,1b1}$ of neat water as a reference value. Any possible changes of VIE$_{vac,1b1}$ in a solution or the aqueous e$\Phi$ are disregarded in this approach. **B)** A bias applied to the LJ shifts all liquid features under the influence of an accelerating field, E$_{acc}$ (blue spectrum); the gas-phase PE signal is smeared out and does not appear here. Biasing reveals the full LET and cutoff energy of the sample spectrum, E$_{cut(s)}$. Without bias (grey spectrum), the real cutoff is obscured by the work-function difference between the liquid and analyzer, $\Delta e\Phi$, and one would instead measure a setup-dependent cutoff energy, E$_{cut(A)}$. E$_{cut(s)}$ constitutes a low-energy limit for photoelectrons to still overcome the liquid-surface barrier, and is thus connected to the local vacuum level above the LJ surface, E$_v^{loc}$. The precisely known photon energy hν (vertical purple arrow) is used to map E$_v^{loc}$ onto the measured spectrum and define the VIE$_{vac}$ scale. Note that in general E$_v^{loc}$ will deviate from E$_v^\infty$ due to the intrinsic surface potential $\chi^d$ / e$\varphi_{outer}$ (see panel A and the text for details). Any extrinsic potentials are irrelevant in the applied bias case because the only relevant quantity is the energetic separation of the PE features from E$_{cut}$, $\Delta E_w$ (blue arrow). **C)** As for B) but for an arbitrary aqueous solution; here, the spectra are arbitrarily aligned to the cutoff, which at the same time aligns E$_v^{loc}$. Changes in ΔE_w directly translate to changes in the VIE. The lower part of this panel shows the full unbiased spectrum (compare to the spectra shown in the inset in panel A and bottom part of panel B). **D)** The liquid water spectrum (dark blue) is energy-referenced to a common Fermi level, $E_F$, which defines the ionization energy scale with respect to Fermi, $VIE_{EF}$. This is achieved by separately measuring a metallic sample (red spectrum) in electrical contact and equilibrium with the liquid. The liquid-phase measurements must be performed with a sufficient amount of dissolved electrolyte to suppress the streaming potential and assure good conductivity. The Fermi-alignment with the apparatus leads to an offset of the local vacuum potentials as shown in the top inset in panel D. This creates an intrinsic potential difference due to the generally different $e\Phi$ values between the sample and the apparatus (detector). Thus, the measurement is usually not performed under field-free conditions, unlike Method 1. The difference between the $VIE_{vac}$ and $VIE_{EF}$ scales yields water's work function, $e\Phi_{water}$. We additionally sketch (light blue), the situation where $e\Phi$ changes and the valence spectrum shifts with respect to $E_F$ upon build-up of a surface dipole arising from adsorbed interfacial anions and cations (here, representative of a surface-active TBAI aqueous solution; although this latter detail is not depicted). $TBAI_{(aq)}$ is known to exhibit a pronounced surface-dipole layer comprised of spatially slightly spatially separated maxima in the $TBA^+$ and $I^-$ concentration profiles,[86] which may lead to a reduction in $e\Phi$. This in turn would shift the position of $E_v^{loc}$ of the TBAI solution with respect to the metallic sample.

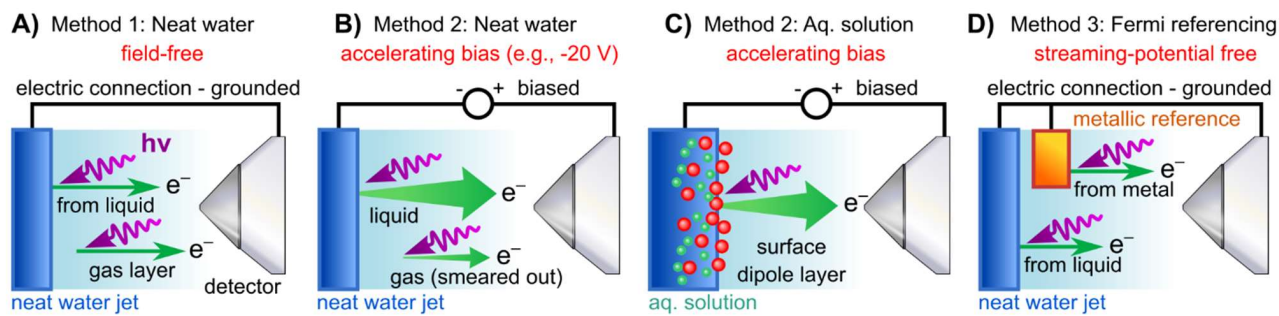

**Figure 2**: Schematic setups for the measurement procedures introduced in Figure 1. A) Electrically grounded (nearly) neat water LJ with a precisely tuned salt concentration to achieve a field-free condition for gas-phase referencing. B) Negatively biased LJ used to reveal $E_{cut}$ in the liquid spectrum for energy referencing; gas- and liquid PE contributions are energetically separated in the field gradient. C) Same as B) but for an aqueous solution (here, featuring a surface-active solute). Changes in VIEs can be directly observed. D) Similar to A) but with the addition of a metallic reference sample held in electrical contact to and mounted within the vicinity of the LJ. The liquid water spectrum can be referenced to the Fermi edge of a metal sample under field-free conditions. Note that the metallic reference sample surface is probed separately from the LJ in the experiments reported here, and thus is not directly affected by any changes at the surface of the solution.

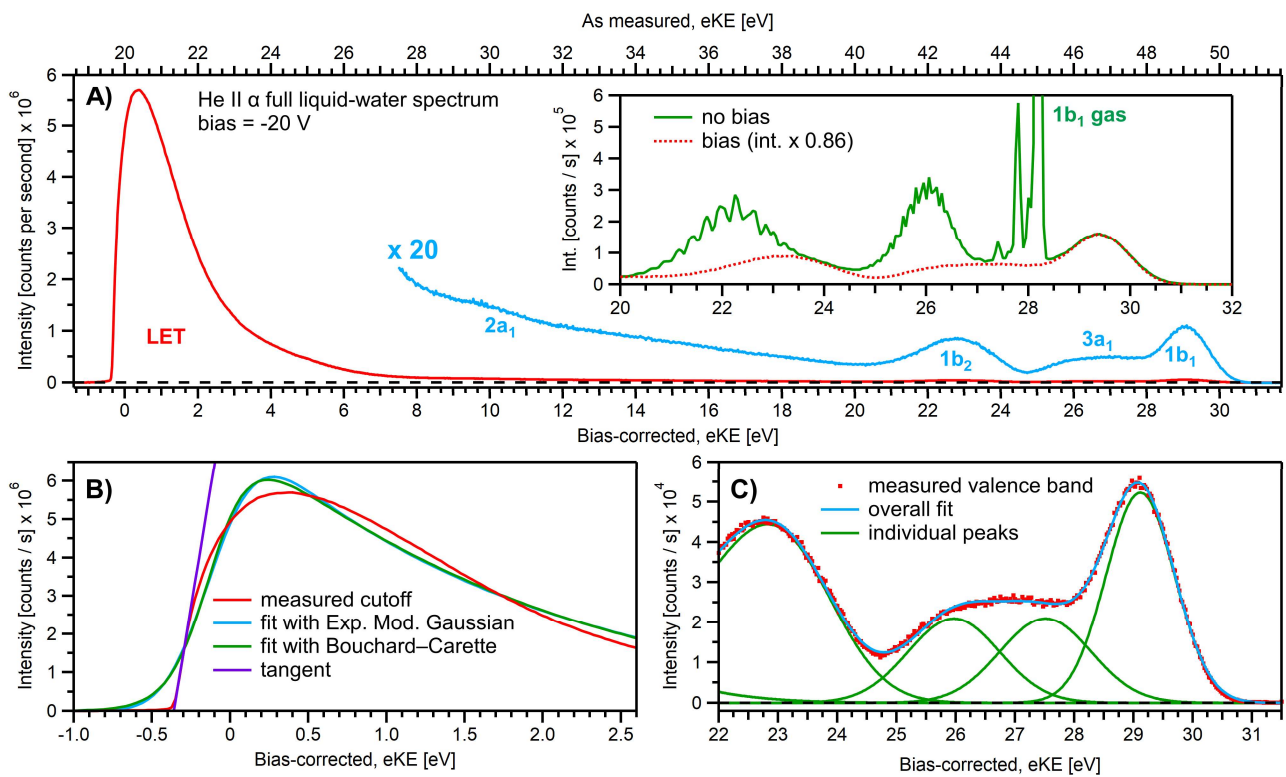

**Figure 3: A)** A representative PE spectrum of liquid water (with 50 mM NaCl added), measured with a monochromatized He II α emission light source (hν = 40.814 eV). Exemplary associated electron count-rates are presented, as reported by the analyzer measurement software. Note that the count rate calibration is that provided by the analyzer manufacturer, which has not been verified under the acquisition conditions implemented here, and correspondingly should be considered a coarse guide to the overall experimental acquisition conditions only. A bias voltage of -20 V was applied to separate the liquid- and gas-phase contributions as well as to expose the low-KE tail (LET) region. The as-measured eKE is shown on the top x-axis in A), with the bias-corrected scale shown on the lower x-axis. The same spectrum with the intensity multiplied by 20 shows the full valence band of water. The inset compares the valence region with and without an applied bias, exposing the gas-phase contribution. **B)** A close-up of the cutoff region with three analysis methods applied as described in the main body of the text. The bias-corrected x-axis scale is plotted and the residual gas-phase contribution has been subtracted. **C)** A close-up on the valence spectral region with a cumulative Gaussian fit to all ionization peaks/molecular orbital contributions (only the three highest orbitals are visible here), also plotted on the bias-corrected x-axis scale.

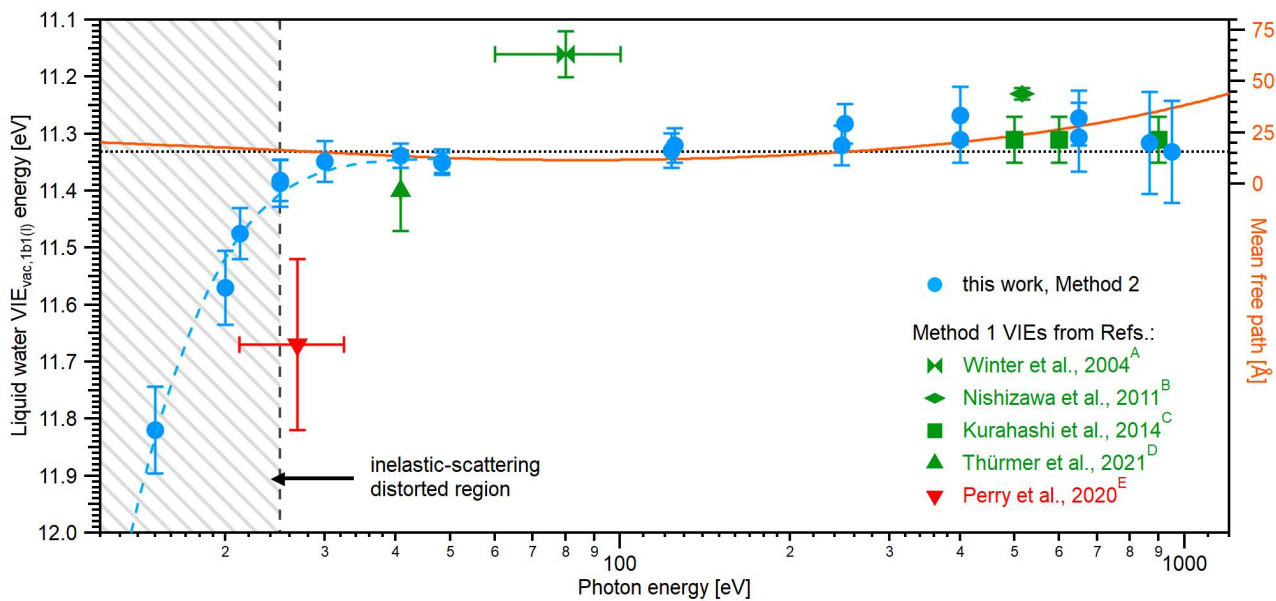

**Figure 4:** An overview of the determined $VIE_{vac,1b1(l)}$ values as a function of photon energy. Green squares show results obtained with the gas-phase referencing method, Method 1, where the field-free condition was achieved by carefully compensating for all potentials with a specific salt concentration: A) from Ref. [4], B) from Ref. [27], C) from Refs. [28] (where the used photon energies have been confirmed by the authors[93]), and D) from Ref. [84]. The value E) associated with the red triangle was instead obtained by applying a compensation bias voltage between the detection system and LJ to achieve a field-free condition.[29] The values determined in this work, using Method 2, are shown as dark blue circles. Note that the $VIE_{vac,1b1(l)}$ values *seemingly* shift to higher values at lower photon energies, which corresponds to low eKEs for the lowest ionization energy ($1b_1$) photoelectrons (blue dashed line in the gray hatched area). This is, however, an artifact arising from increased inelastic electron scattering at low eKEs. The averaged, nascent VIE or binding-energy value – minimally affected by electronic scattering effects – is marked with the black dashed line. Error bars show the confidence interval as reported in the studies / resulting from the analysis of our data. The electron mean free path from Ref. [85] is shown as a guide to the eye in orange and on the scale to the right. While we cannot distinguish any depth dependence to $VIE_{vac,1b1(l)}$ with the current error bars, the possibility of slight changes in $VIE_{vac}$ with depth are discussed in the text.

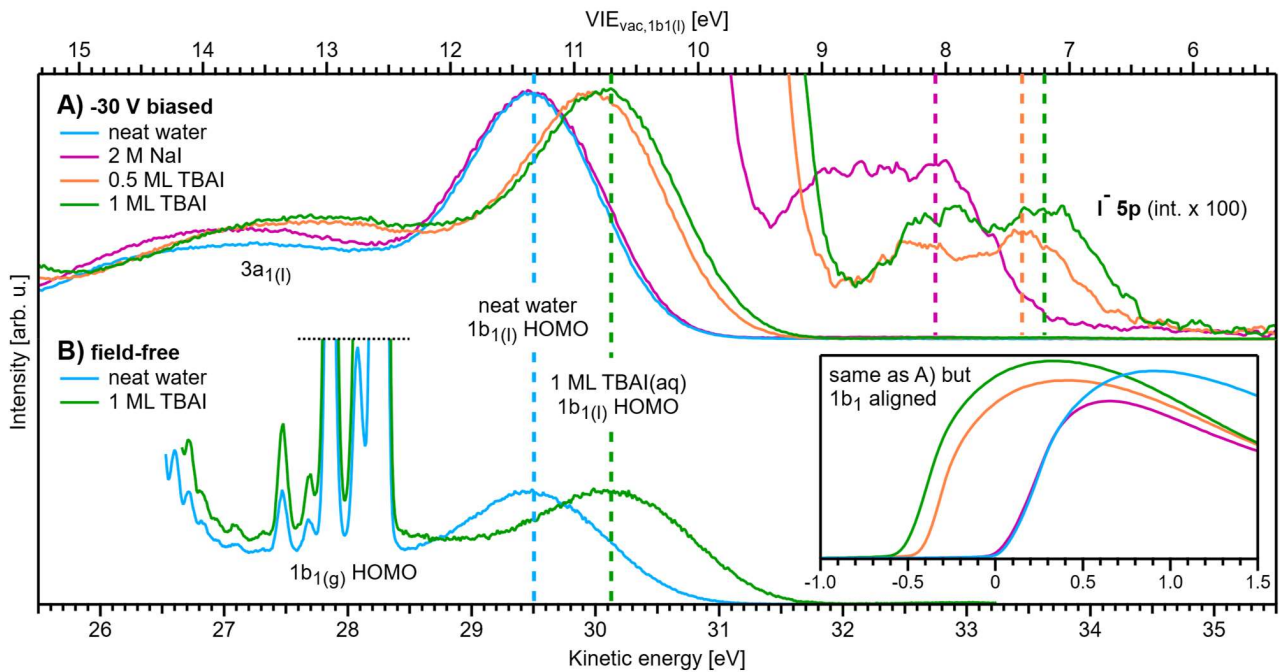

**Figure 5:** Changes in $VIE_{vac,1b1}$ for representative aqueous solutions, both with an applied bias and a grounded jet. All spectra were recorded with He II α emission (hν = 40.814 eV). **A)** Spectra measured with a bias voltage of -30 V. Each cutoff position was then aligned to eKE = 0 eV, which immediately visualizes $VIE_{vac}$ changes as shifts of the liquid $1b_1$ HOMO position; the top axis shows the corresponding $VIE_{vac,1b1(l)}$ energy scale. The bottom-right inset shows the same spectra aligned to the $1b_1$ HOMO position, which instead show a shift in the cutoff position; both presentations are equivalent. Neat water serves as a reference position (dark blue line; about 50 mM NaCl was added here, but the precise value is irrelevant for this method). All spectra are normalized to the same $1b_1$ peak height. The spectra are shown multiplied by a factor of 100 (and smoothed with a 5-point boxcar averaging) to reveal the I⁻ 5p solute feature to the top-right. The position of the $5p_{3/2}$ peak is marked with a dashed line in each case. **B)** Spectra measured with a grounded jet. The salt concentration for the (nearly) neat water spectrum (blue line) was precisely tuned to achieve field-free conditions (2.5 mM NaCl was optimal here). The spectra are aligned so that the $1b_1$ position of neat water is matched with A). The same shift is observed with 1 ML TBAI (green line) as in A, which shows the equivalence of Methods 1 and 2. Here, TBAI aqueous solution serves as a special case, where the field-free condition is preserved even for the solution, which makes a direct comparison possible in the first place. In general, the solutions and delivery conditions generate non-zero extrinsic and intrinsic potentials which impose an unknown additional energy shift to the liquid spectra.

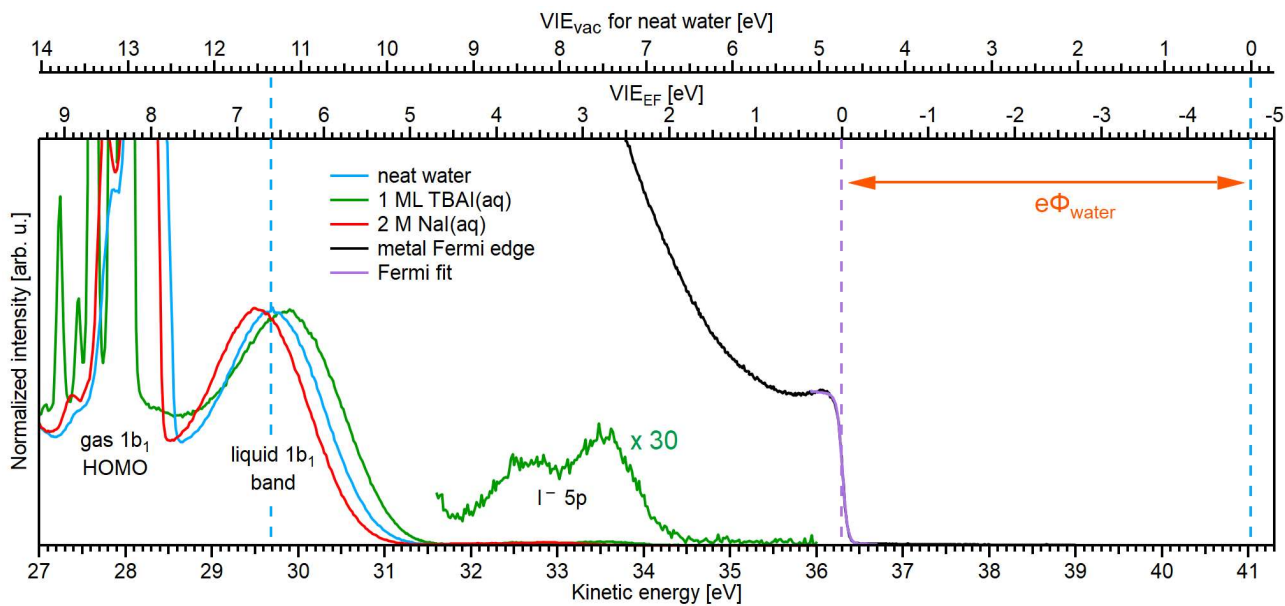

**Figure 6:** Determination of $VIE_{EF,1b1(l)}$ for neat water (blue, with an optimal NaCl concentration of 50 mM; see the main body of the text for details) and the limitations of this method for aqueous solutions, exemplified here for 1 ML $TBAI_{(aq)}$ (green) and 2 M $NaI_{(aq)}$ (red) solutions. The relative energy position of liquid water's lowest energy $1b_1$ ionization feature and the Fermi edge of a metallic reference sample were separately recorded using He IIα emission (hν = 40.814 eV). A sample bias was not applied in either case and the bottom axis shows the as-measured kinetic energy scale of the detector. To the right, the highest eKE feature of the metal spectrum is shown in black (only the Fermi edge is visible). The position and spectral shape of the measured metal spectrum was unchanged following the introduction of the LJ and solution. The Fermi edge was fit with a Fermi function[23] (purple line), and its position defines the zero point of the $VIE_{EF}$ energy scale in the spectrum (lower axis scale at the top of the panel). This enables us to determine the $VIE_{EF,1b1(l)}$ value of 6.60 ± 0.08 eV and a $e\Phi_{water}$ value of 4.73 ± 0.08 eV for (almost) neat water. For the 2 M $NaI_{(aq)}$ solution, the $1b_1$ peak is shifted towards lower eKEs (higher $VIE_{EF}$), which most likely arises from additional extrinsic fields as opposed to a real change of the aqueous electronic structure for this solution (compare to Figure SI-7); the $VIE_{vac}$ values underwent insignificant changes in going from neat water and 2 M $NaI_{(aq)}$ solutions (compare to Figure 5). Without proper assessment of additional potentials, such as the streaming potential or surface charge, it is in principle impossible to accurately reference eKEs to $E_F$ or judge associated changes in $VIE_{EF}$ in this case. In the case of $TBAI_{(aq)}$, on the other hand, the $1b_1$ shifts towards higher eKEs (lower $VIE_{EF}$). It can be argued that this shift is caused by band-bending at the liquid interface (see text for details). Multiplying the $TBAI_{(aq)}$ spectrum by a factor of 30 reveals the $I^-$ 5p solute features around eKE ≈ 33 eV, corresponding to $VIE_{EF}$ values of 3.80 ± 0.10 eV and 2.84 ± 0.10 eV for the $5p_{1/2}$ and $5p_{3/2}$ levels, respectively.

# Tables

| Measured at | hν (eV) | VIE$_{vac,1b1(l)}$ (eV) | VIE$_{vac,O1s(l)}$ (eV) |
|---|---|---|---|
| DESIRS, SOLEIL | 15.00 ± 0.03 | 11.82 ± 0.08 | |
| DESIRS, SOLEIL | 19.99 ± 0.03 | 11.58 ± 0.07 | |
| Laboratory, FHI Berlin | 21.218 ± 0.001 | 11.48 ± 0.05 | |
| DESIRS, SOLEIL | 24.98 ± 0.03 | 11.38 ± 0.04 | |
| **DESIRS, SOLEIL** | **29.97 ± 0.030** | **11.35 ± 0.04** | |
| **Laboratory, FHI Berlin** | **40.813 ± 0.001** | **11.34 ± 0.03** | |
| **Laboratory, FHI Berlin** | **48.372 ± 0.001** | **11.35 ± 0.03** | |
| **U49-2_PGM-1, BESSY II** | **123.464 ± 0.004** | **11.33 ± 0.03** | |
| **U49-2_PGM-1, BESSY II** | **246.927 ± 0.005** | **11.32 ± 0.04** | |
| **P04, PETRA III** | **249.99 ± 0.02** | **11.28 ± 0.04** | |
| **P04, PETRA III** | **400.01 ± 0.03** | **11.31 ± 0.04** | |
| **U49-2_PGM-1, BESSY II** | **400.868 ± 0.004** | **11.27 ± 0.05** | |
| P04, PETRA III | 650.03 ± 0.03 | 11.27 ± 0.05 | 538.08 ± 0.05 |
| U49-2_PGM-1, BESSY II | 649.67 ± 0.03 | 11.31 ± 0.06 | 538.13 ± 0.05 |
| U49-2_PGM-1, BESSY II | 867.29 ± 0.01 | 11.32 ± 0.09 | 538.07 ± 0.07 |
| P04, PETRA III | 950.06 ± 0.03 | 11.33 ± 0.09 | 538.04 ± 0.08 |

**Table 1:** VIE$_{vac,1b1(l)}$ and VIE$_{vac,O1s(l)}$ values of the liquid water valence 1b$_1$ band and O 1s core-level peaks, respectively. The values were extracted from the spectra measured at different photon energies using the absolute referencing analysis method, Method 2. These values represent the averages of all measurements performed at the respective photon energy. The values in bold font are deemed to be essentially free of electron scattering based distortions of the measured VIE$_{vac}$ values, while still being minimally affected by spectral distortions associated with the applied bias. The VIE$_{vac}$ values shown in bold font can alternatively be referenced to the Fermi level, VIE$_{EF}$. Such values can be ascertained by subtracting the work function of liquid water, eΦ$_{water}$, determined here from the VIE$_{vac}$ values. See the main body of the text for further details.

# Notes and References

N1  Note that the term 'band' for the assignment of 'spectral bands' in molecular spectroscopy has a different meaning to that applied within the band-structure context of condensed matter.

N2  Although the importance of the determination of the cutoff energy in liquid-jet photoelectron spectra, with the aim of quantifying work functions from aqueous solutions, has been accented already in 2003,[94] this approach was barely further considered for the subsequent 10-15 years. Arguably, the reason is a combination of the gas-phase-references being such an easy and convenient (although problematic) method, and technical realization of low-energy electron detection with liquid-jet PES setups. For a long time, and this is still true in many cases, LJ-PES experiments with HEAs were barely designed to detect low-kinetic energy electrons, typically due to insufficient magnetic shielding and likely because it had yet to be demonstrated that liquid phase PES is capable of accessing characteristic condensed-matter properties. Finally, the ability to properly apply a bias voltage to a liquid jet had to be thoroughly explored, an issue with remaining open questions, such as the degree of the deleterious effects of biasing an entire sample delivery assembly, as opposed to just the liquid stream.

N3  We will refer to the inelastic scattering tail as the low-KE tail or LET curve throughout the manuscript, in contrast to the often-used term secondary electron energy distribution (SEED) curve described in previous studies. This is consistent with the fact that at lower photon energies, when the kinetic energy of the primary electron is too low for efficient secondary electron generation via, *e.g.*, impact ionization, the inelastic scattering background is not fully comprised of secondary electrons. Thus, the term SEED cannot be used for aqueous solution spectra recorded at ~20 eV photon energies and below.[30] The term LET is adopted to avoid misleading connotations about the origin of this low energy signal.

N4  In fact, it is possible to deliberately modify $e\Phi$, *e.g.*, by adsorption of molecules on the sample surface, which typically induces or alters a pre-existing surface dipole, the associated value of $\chi^d$, and necessarily the value of $E_v^{loc}$.[24, 35, 91] This creates or modifies the energetic barrier for the photoelectrons escaping from the sample into vacuum, and can be detected as a change in KE of the emitted electrons. For a sample with a truly uncharged, amorphous, apolar surface, we note that $E_v^\infty = E_F + \bar{\mu} = E_F + e\Phi$. However, if an intrinsic dipolar surface potential exists, the first equality holds, and the second generally will not. In such a case, and depending on the geometry of the liquid surface and its overlap with an ionizing light source, condensed phase VIEs or binding energies will be offset by an experimental-geometry-averaged amount with respect to $E_v^\infty$ due to the average offset of $E_v^{loc}$.

N5  Here we contrast the electronic structure of liquid water with that of metallic, ionic, or covalent macroscopic solids, where delocalized electronic states are formed via atomic valence energy level interactions that generate quasi-continua of energy levels, termed bands. An important consequence of this behavior is that PES spectra recorded from non-molecular systems generally exhibit broad valence features that elude association with specific VIEs. Thus, VIE is a quantity less typically encountered in a condensed-matter electronic structure context, with band edge electronic structure descriptors more commonly being reported. It is of great interest to explore how these different descriptors and experimental observables interconnect within a unified 'band structure' description of typical solids, liquid water, and aqueous solutions, although this is beyond the scope of the present study.

N6    The problem of ionization-induced charging is well-known in solid insulator studies and is usually sufficiently counteracted using neutralization instrumentation such as electron flood guns.[95] Notably, the charging of the surface of a volatile, flowing aqueous solution in a low-vacuum environment cannot be compensated in this way.

N7    The transmission function of the HEA generally influences the relative signal intensities over larger energy ranges, and especially at very small eKEs, the electron signal is distorted as slow electrons are particularly affected by stray fields (which is another reason to apply an accelerating bias). This makes it difficult to compare exact relative intensities over an energy range larger than about 30-40 eV, something which is beyond the scope of the findings presented here. Any feature within a smaller energy window, such as the valence band region or the cutoff region can be separately analyzed without further correction, since the transmission function will vary minimally over such a small energy range (assuming the cutoff electrons are sufficiently accelerated by an applied bias). A particularly important aspect is the potential effect of the analyzer electron transmission function on the LET shape upon application of a bias voltage. We find, however, that this effect has a negligibly small impact on the value of the extracted absolute VIE values, as detailed in Figure SI-3.

N8    A tightly focused ionizing beam – such as those provided by synchrotron or laser light sources – primarily probes gaseous molecules in the immediate vicinity of the LJ surface. The associated spectra may be only mildly energetically broadened in a field gradient and the relatively low potential difference spanning the probed volume. Consequently, associated measurements may present an apparently sharp gas-phase PE signal, despite the presence of an extrinsic field gradient between the sample and electron analyzer.

N9    Such a measurement would be forced to deal with a further complication: The electrons from the metal would experience both the metal-solution interfacial potential and the aqueous-vacuum potential, whereas the solution phase electrons would experience the latter only. It can still be argued, however, that $E_F$ would be equilibrated throughout this system as long as the solution was sufficiently conductive. Considering alternative methodologies for the co-determination of solution- and solid-phase electron energetics, application of the 'dip-and-pull' PES method may seem appropriate.[96] However, a significant associated challenge lies in achieving sufficient control over the composition and cleanliness of the solution – vacuum interface, as well as the composition of the solution bulk following solution pulling and under a significant cumulative ionizing radiation load.

N10    Note that an optimal concentration of 30 mM (at a flow rate of 0.5 ml/min and at room temperature) has also been reported to establish field-free condtions,[28] which however depends on experimental parameters like the size and sign of the sample-spectrometer contact potential or work function difference, $\Delta e\Phi$. We remind the reader that our field-free conditions were established under rather different conditions with a 2.5 mM NaI concentration, a flow-rate of 0.8 ml/min, and a tapered fused silica capillary nozzle with a 30 μm orifice diameter.

N11    Such a nonlinear shift in the streaming potential has been observed before. However, no explanation has been given for the high-concentration behavior.[28]

# Supporting Information

## Accurate Vertical Ionization Energy and Work Function Determinations of Liquid Water and Aqueous Solutions

Stephan Thürmer,[1]* Sebastian Malerz,[2] Florian Trinter,[2,3] Uwe Hergenhahn,[2] Chin Lee,[4,5] Daniel M. Neumark,[4,5] Gerard Meijer,[2] Bernd Winter,[2]* and Iain Wilkinson[6]*

[1] Department of Chemistry, Graduate School of Science, Kyoto University, Kitashirakawa-Oiwakecho, Sakyo-Ku, Kyoto 606-8502, Japan
[2] Molecular Physics Department, Fritz-Haber-Institut der Max-Planck-Gesellschaft, Faradayweg 4-6, 14195 Berlin, Germany
[3] Institut für Kernphysik, Goethe-Universität, Max-von-Laue-Straße 1, 60438 Frankfurt am Main, Germany
[4] Chemical Sciences Division, Lawrence Berkeley National Laboratory, Berkeley, CA, 94720 USA
[5] Department of Chemistry, University of California, Berkeley, CA, 94720 USA
[6] Department of Locally-Sensitive & Time-Resolved Spectroscopy, Helmholtz-Zentrum Berlin für Materialien und Energie, Hahn-Meitner-Platz 1, 14109 Berlin, Germany

# List of Symbols and Acronyms

**General:** subscript (l) = liquid-phase features, subscript (g) = gas-phase features, subscript (aq) = solute features in an aqueous solution, subscript (sol) = water features in presence of a solute.

| Symbol / Acronym | Meaning |
|---|---|
| PE | photoelectron |
| PES | photoelectron spectroscopy |
| HEA | hemispherical electron energy analyzer |
| ToF | time-of-flight spectrometer |
| LJ | liquid jet |
| mM / M | millimolar / molar |
| ML | monolayer |
| BB | band-bending |
| HOMO | highest occupied molecular orbital |
| LET | low-kinetic energy tail |
| SEED | secondary electron energy distribution |
| IMFP | inelastic mean free path |
| HHG | high harmonic generation |
| VUV / EUV | vacuum-ultraviolet / extreme ultraviolet |
| $h\nu$ / $E_{ph}$ | photon energy |
| eKE / KE | (electron) kinetic energy |
| $eKE_{(meas)}$ | as-measured electron kinetic energy |
| VIE | vertical ionization energy |
| (V)BE | (vertical) binding energy |
| $VIE_{vac}$ | VIE with respect to the vacuum energy level |
| $VIE_{EF}$ | VIE with respect to the Fermi level |
| $VIE_{EF,1b1}$ | VIE of neat water's $1b_{1peak}$ with respect to the Fermi level |
| $VIE_{EF,solute}$ | VIE of the [solute] peak with respect to the Fermi level |
| $VIE_{vac,1b1}$ | VIE of neat water's $1b_{1peak}$ with respect to the vacuum level |
| $VIE_{vac,solute}$ | VIE of the [solute] peak with respect to the vacuum level |
| $VIE_{vac,O1s}$ | VIE of the O 1s core-level for neat water with respect to the vacuum level |
| $E_{cut(s)}$ | real cutoff energy of the sample spectrum |
| $E_{cut(A)}$ | setup / analyzer-dependent cutoff energy |
| $\Delta E_{g-l}$ | energy distance between gas and liquid peaks |
| $\Delta E_{l-l}$ | energy distance between liquid solute and solvent peaks |
| $\Delta E_w$ | energetic width of the spectrum = distance between cutoff and peak features in neat water |
| $\Delta E_{w(sol)}$ | energetic width of the spectrum for an aqueous solution |
| $E_v^\infty$ | theoretical vacuum energy level at infinity, far away from any matter |
| $E_v^{loc}$ | local vacuum energy level above the liquid's surface |
| $E_v^{det}$ | local vacuum energy level of the detector / experimental setup |
| $E_F$ | Fermi energy or Fermi level |
| $e\Phi_{water}$ | work function of neat liquid water |
| $e\Phi_{sol}$ | work function of an aqueous solution |
| $e\Phi_A$ | work function of the analyzer / experimental setup |
| $\Delta e\Phi$ | work function difference / contact potential between sample and experimental setup |
| $\Phi_{str}$ | streaming potential of a flowing liquid |
| $I_{str}$ | streaming current of a flowing liquid |
| $\mu$ | chemical potential; equal to $E_F$ |
| $\bar{\mu}$ | electrochemical potential |
| $\phi_{s,jet}$ | total extrinsic surface potential of the liquid jet |
| $\chi^d$ / $e\varphi_{outer}$ | intrinsic interfacial dipole potential / outer (Volta) potential |
| $S_{distort}$ | bias-voltage-induced energy-distortion scaling factor |

## 1 Useful Considerations on the Presentation of Electron Binding Energies in PES

We re-evaluate what should be the most useful energy-axis presentation used in condensed-phase PE spectroscopy, as recently discussed in the context of core-level studies.[1] This appears to be timely, partially triggered by specific phenomena encountered in aqueous-solution measurements. If we consider PE spectroscopy as a reflection of the photoelectric effect,[2] implying the detection of a direct photoelectron, the commonly applied VIE (or equivalently binding energy, BE) axis (with reference to $E_v$ or $E_F$, depending on context) has been advised, dating back to the early days of PE spectroscopy and practiced in text books.[3] Here, one makes use of the (simplified) relationship: VIE = KE – hν. Although convenient, this common practice is correct only if the spectra have no dependence on photon energy. When final-state effects (such as photoionized-state vibrational progressions in molecular spectra) are considered, a direct correspondence between KE and VIE breaks down. In molecular spectra, additional axes or markers are often introduced to indicate peak structures originating from the same electronic configuration. In a strict sense, values should thus be reported as KEs, while at the same time the peaks in this progression can be shown on a VIE scale. Similarly, in X-ray PE spectra, the occurrence of additional signals originating from autoionization processes no longer justifies the application of a VIE scale, as such features most often have a fixed KE (which is measured in the experiment) due to their very nature and become photon-energy-dependent when displayed on a VIE scale.[4]

Arguably, the applicability of a VIE scale is seriously questionable when presenting PE spectra from condensed-phase samples. Here, there are a plethora of final-state effects, starting from satellite features due to plasmon loss, (charge-transfer) multiplets, or screening, as well as peak-skewing mechanisms due to, *e.g.*, electron-hole interactions, and resonantly enhanced features that only appear at certain photon energies.[3] These extra signal contributions are incidentally assigned a "meaningless" VIE value when such a scale is globally adopted, and have to be marked and painstakingly discussed. While such discussions can be accommodated, the benefits of assigning a global VIE scale widely disappear. For instance, incidental assignment of a VIE scale to inelastically scattered electrons is incorrect *per se*. This problem is exacerbated when the low KE tail, LET, signal exhibits peak-like enhancements, as observed for the liquid-water case.[5] We thus recommend a more rigorous way of displaying energy scales of PE spectra, as indicated in Figs. 1, 5 and 6 in the main text. PE spectra should be plotted on an absolute or relative KE scale which may be corrected for experiment-specific effects. Then, a second energy scale is introduced which marks calibrated ionization or binding energies of relevant PE features. Additional scales may be employed to mark vibrational progressions or shifted features, *e.g.*, satellite features at a constant energy offset. Indeed, other data processing steps may be required in advance of the proposed procedure. For example, spectra measured using ToF spectrometers have to be converted from flight time to kinetic energy scales with assumptions or calibrations made for the ionization time zero and the particular geometrical arrangement of the flight tube. Even HEA spectrometers have to be pre-calibrated using known gaseous PE features to measure the correct kinetic energy, which remains an imperfect process, with even small changes in sample placement or other changes to the experimental setup potentially slightly altering the measured eKE. In practice, associated corrections of the eKE scale may be applied to the data during analysis.

## 2 Extrinsic potentials occurring with liquid jets

Here, we briefly comment on the various sources of extrinsic potentials occurring in experiments with free-flowing liquid jets (LJs). There are three major components: the photoionization-induced surface potential (arising from charge-buildup due to water's insulating properties), the contact potential between the liquid and the detector ΔΦ, and the streaming potential (which has its origin in kinematic charging of the injected LJ) all of which lead to an electric field between the surface of the jet and the grounded analyzer orifice.[5-7] Often, extrinsic surface potential and streaming potential terms are used interchangeably in the LJ-PES literature, as no experimental distinction is possible; here we use the extrinsic surface-potential term to describe the potential associated with external parameters, $\phi_{s,jet}$. The magnitude of the three aforementioned surface-potential contributions can be influenced by adding an electrolyte to the solution. Notably, a sufficient amount of electrolyte suppresses net jet charging by increasing the conductivity of the liquid, but an overdosing of electrolyte may lead to non-zero streaming potentials.[7-9] A temperature and flow-rate dependence has been additionally identified.[8,10] The resulting electric field between the jet and the analyzer will change if $\phi_{s,jet}$ changes, and exactly this dependence has been used to track *changes* of $\phi_{s,jet}$, as inferred from the energy shifts of $VIE_{vac,1b1(g)}$ (or $\Delta E_{g-l} = VIE_{vac,1b1(g)} - VIE_{vac,1b1(l)}$). We note that the resulting effect, *i.e.*, an energy shift of the gas-phase peaks *in the direct vicinity* of the liquid surface, is indistinguishable from shifts caused by differences in $e\Phi$ between the LJ and the analyzer. Establishing a *field-free* condition for the gas-phase referencing in Method 1 thus implies a compensation to zero of *all* these effects combined (with possible associated modification of the intrinsic surface-dipole potential). The so far common procedure to achieve this is to tune the amount of electrolyte (usually NaCl or NaI) until a (on average) field-free condition has been reached between the sample and analyzer. This condition was referred to as streaming-potential compensation,[7] but one may argue that instead a *residual* streaming potential is engineered to exactly compensate other effects such as those originating from contact potential / $e\Phi$ differences. We explicitly determine different electrolyte concentrations to achieve *field-free conditions* when applying Method 1 (which was 2.5 mM at room temperature and a flow rate of 0.8 ml/min.) or to achieve *streaming-potential-free* conditions when applying Method 3 (30-50 mM).

## 3 Accurate determination of ionizing photon energies

We consider methods for the accurate determination of ionizing photon energy ($E_{ph}$), which is an essential step for the proper implementation of energy referencing Method 2. Our ionizing-radiation source considerations span the most commonly adopted sources: discharge lamps, synchrotron radiation beamlines, and laser-based high harmonic generation (HHG) setups. In this regard, discharge lamp photon energies are intrinsically precisely determined by their sharp atomic transitions. In the case of our helium discharge lamp, a simple grating monochromator, placed between the plasma-discharge chamber and the exit capillary, is used to select a single, desired emission line; each emission line has an intrinsic bandwidth of just 1-2 meV around the associated transition energy. This makes these sources ideal for calibration measurements, provided the photon energy is high enough to prevent distortion of PE features by inelastic scattering at low kinetic energies (KEs).[11]

Synchrotron beamlines, on the other hand, select a narrow energy window from a rather broad radiation source, which emits either over a wide ('white' light in case of single bending magnets) or narrower, but continuously tunable (wigglers and undulators) spectral range. Both the radiation source and the beamline work in tandem to output the desired photon energy. A multitude of influences, such as mechanical offsets, thermal expansion, aberrations, changes in mirror illumination, and slit settings need to be precisely controlled and checked to assure the correct photon energy output. For this reason, we recommend to measure the photon energy at the end of the beamline at least on the same day, and ideally directly before and/or after the VIE calibration measurement. For this, a dedicated apparatus incorporated within (and towards the end of) the beamline (*e.g.*, gas-phase ionization cell, calibrated X-ray spectrometer etc.) can be used, but it is often more practical to use the photoelectron spectrometer itself. Here two approaches can be adopted. First, the spectrum of the same PE feature can be recorded with both the first and second harmonic outputs of the beamline. Second, the (eKE-integrated) partial electron yield photoabsorption spectrum can be recorded via a short photon-energy scan across a precisely known gas-phase atomic or molecular transition. Both methods were adopted to photon-energy-calibrate the soft X-ray VIE results (hv = 124 – 950 eV) reported in Figure 4 and Table 1 in the main body of the text (in specific hv cases, these methods were sequentially implemented to doubly-calibrate the soft X-ray energies and cross-check the consistency of the two photon-energy calibration methods). The first approach exploits the fact that any grating monochromator will direct several diffraction orders (hv, 2hv, 3hv etc.) along the beamline at a specific nominal photon energy setting, generally at successively reduced efficiencies for the higher diffraction orders. Thus, each PE feature will appear at multiple KEs and the measurement of both the first and second order PE signals makes it possible to calculate $E_{ph}$ from the energetic separation of the two signals, $E_{ph} = KE_{2nd} - KE_{1st}$ (provided the spectrometer eKE axis is sufficiently linear). This approach has the drawback that higher-order light is often actively suppressed at modern synchrotron radiation beamlines and both the ionization cross sections and spectrometer transmission quickly diminish at higher KEs, generally leading to very weak or even unmeasurable second-order PE signals. Furthermore, broad or noisy spectral features measured with low-flux, residual higher-harmonic sources have the potential to introduce considerable errors in the determination of energetic distances. Also, using liquid-phase features (*e.g.*, the water O 1s core level or solute peaks) for this calibration purpose requires sufficiently equilibrated and stable conditions between the first and second harmonic measurements, as slight energetic shifts caused by changing surface potentials lead to errors in the determination of $E_{ph}$. The second approach of scanning the photon energy over a known gas-phase resonant transition feature usually does not suffer from such signal-intensity problems. Here, the partial electron yield PE spectra are generally recorded and are subsequently integrated over the eKE axis to produce a proxy for the true X-ray absorption spectrum. The resonances are identified as PE signal enhancements on the $E_{ph}$ energy axis. In the soft X-ray measurements reported in the main body of the text, the $Ar_{(g)}$ $2p_{3/2} \rightarrow$ 3d transition at 246.927 ± 0.001 eV,[12] $CO_{(g)}$ C 1s → π* (v=0) transition at 287.41 ± 0.005 eV,[13] $N_{2(g)}$ N 1s → $π_g$* (v=0) transition at 400.868 ± 0.001 eV,[14] $CO_{(g)}$ O 1s → π* transition at 534.21 ± 0.09 eV,[15] and $Ne_{(g)}$ 1s → 3p transition at 867.29 ± 0.01 eV[16] were implemented to calibrate the photon energies. Resonant photon energies of 123.464 ± 0.001 eV (2nd harmonic of the beamline was resonant with the $Ar_{(g)}$ $2p_{3/2} \rightarrow$ 3d transition), 246.927 ± 0.001 eV, 400.868 ± 0.001 eV, and 867.29 ± 0.01 eV were accordingly set for PES measurements and the cumulative resonant absorption data was used to produce overall beamline calibrations that could be used to precisely determine the nominally set 650 eV and 950 eV photon energies in addition. Generally, the resonant absorption method has the drawback of necessitating measurements at (or close to) suitably calibrated atomic or molecular transition energies. Notably, it is not possible to accurately determine the photon energy by measuring a single PE spectrum of a known peak alone, as this would not correct for any intrinsic energy offsets in the measured kinetic energy scale of the spectrometer, which cannot be disentangled from offsets in the photon energy, *i.e.*, $KE_{meas} = (E_{ph} + \Delta KE_{error}) - IE_{ref}$.

For laser-based HHG sources, the photon energy notably sensitively depends on the driving laser and gas-cell parameters at the point of harmonic generation,[17-19] as well as the subsequent monochromatizing beamline parameters. Careful, at least daily, photon energy calibrations as well as source-stabilization measures are accordingly generally required to make full use of Method 2 with such sources. With ToF-based spectrometers – which are most often implemented with photon-number-limited, monochromatized (but still relatively broadband) HHG sources – the laser pointing and laser-liquid-jet crossing position in front of the spectrometer entrance aperture has the further effect of impacting the spectrometer eKE calibration. To enable both accurate hv and eKE calibration, the following procedure is suggested. With hv sufficiently in excess of the water or aqueous solution IEs of interest (to avoid deleterious inelastic-scattering-induced effects, *i.e.*, hv ≥ 30 eV), the spectrometer ionization time-zero can often be precisely determined via the reflection of the ionizing radiation towards the electron detector, due to its additional sensitivity to EUV photons. Conveniently, the Fresnel reflection of the HHG beam from a liquid microjet placed in front of a ToF-electron analyzer is sufficient to generate a significant secondary electron signal at commonly implemented micro-channel plate detectors. The photon arrival time at the LJ and time-zero for the photoemission experiment can therefore be readily calculated based on the measured reflected photon-pulse arrival time at the detector and known or determinable distance-of-flight from the LJ to the detection plane. By subsequently recording gas-phase photoelectron spectra from gas-phase PES peaks – *e.g.*, He 1s, Ne 2p, Ar 3s/3p, Kr 4s/4p, and/or Xe 5p – with well-known VIEs under field-free conditions, the photon energy and field-free spectrometer calibration factor(s) can be precisely determined prior to the absolute liquid-phase VIE measurement using Method 2. Upon switching to a liquid-phase sample, bias-free spectra can be recorded and utilized to subsequently calibrate spectra recorded from a sufficiently negatively biased (and electrically conductive) LJ. Method 2 can then be applied as described in the main body of the text; allowing the spectrometer to equilibrate before recording the LET spectrum, the spectral features of interest, and then the LET spectrum again (to ensure self-consistency of the measurements). To engender the most reliable energy referencing results, we further recommend to cross-check the photon energy calibration with further gas-phase measurements following the liquid-phase experiments.

## 4 Challenges in measuring detector-transmission corrected spectra from liquid targets

Obtaining absolute intensities from photoelectron spectra is a notoriously difficult task, due to the many different factors affecting the detected count rate. One of these factors is the detection systems ability to register an electron of a given kinetic energy (KE), in short, the energy-dependent electron transmission. This response can be calibrated under certain restrictions, e.g., by using a well-characterized and precisely prepared solid or gaseous reference target, or by using an electron gun over a limited energy range (here, smaller KEs are usually more prone to variations by stray fields). In a hemispherical electron analyzer (HEA) several spectrometer characteristics can affect the intensity output (see e.g., Ref. [20]): Non-linear counting at the detector surface, secondary-electron generation within the detection system, and transmission characteristics of the lens and hemispherical parts of the analyzer (sometimes the latter part alone is referred to as the 'transmission function'). Usually, the characterization has to be repeated for different settings of the HEA, such as the pass energy, different lens apertures, hemisphere entrance slit size etc.

However, in contrast to the aforementioned transmission curve characterizations, which, once done, just have to be applied to measured spectra thereafter, volatile liquid microjets pose several significant additional challenges. A primary factor is the, yet-to-be-quantitatively-determined electron KE-dependent inelastic scattering within the vapor layer above the probed liquid surface. This may lead to energy-dependent attenuation and an additional secondary-electron signal contribution. Consequently, energy-, sample-morphology-, and instrument-dependent modulations of the electron transmission function are expected when a volatile liquid jet sample is implemented. This effect will notably be most severe when flat liquid microjets with surface dimensions of the order of the HEA entrance aperture are utilized, with correspondingly high evaporated gas loads. Although, a benefit of such sample morphologies is a relatively low-sensitivity of the transmission function of the analyzer to the sample position, at least in the plane orthogonal to the electron-collection axis.

For experiments utilizing narrow-diameter cylindrical microjets with lower gas load (like those utilized in this work), a further challenge arises in that the HEA transmission function will sensitively depend on the 3D position of the liquid microjet sample with respect to the HEA entrance aperture. In this case, electrons originate from only a very small target region (a μm-radius, curved liquid surface) and slight misalignment of the liquid jet with respect to the HEA's center axis leads to electron KE-dependent signal intensity variations. Another challenge is encountered due to the tendency of liquid jets to create electric potentials between the liquid surface and the HEA entrance aperture. This is further complicated by the changing surface potentials that arise as evaporated species gradually adsorb at the inner surface of the sample chamber and the HEA. As the associated fields will be absent or altered when gas- or solid-phase transmission calibrants are implemented, associated calibrations will generally be inapplicable to the liquid-jet measurements.

Finally, depending on the pumping capacity of the HEA, it may be the case that the electron detector (MCP) is operated at elevated pressure when a liquid microjet is implemented. While this is not the case when the efficiently differentially-pumped HEAs used in this work are implemented, elevated analyzer pressures caused by the gas load associated with the volatile target can affect the amplification efficiency of electron detectors. Collectively, these issues make the calibration of the electron transmission function from a liquid jet target an extremely challenging, and potentially impossible, task.

## 5 Determination procedure for the position of $E_{cut}$ via a tangent

To determine the exact zero-crossing of the spectrum's LET region, the following procedure is employed, as illustrated in Figure SI-4A. The derivative of the data is computed, which peaks at the inflection point of the rising slope of the cutoff. In some cases, a double-structure may arise in the cutoff spectrum, which can originate from variations in the relative alignment of the ionizing source, LJ, and analyzer; here, we always use the lowest-energy peak in the derivative, which gives a consistent $E_{cut}$ value (see Figure SI-4B). If the data is particularly noisy, it can become difficult to reliably extract $E_{cut}$, since noise spikes may obscure the real derivative peak structure and can lead to small additional errors when determining the tangent anchor point. However, slight smoothing of the derivative (as opposed to the raw data) has proven to give a consistent result if the signal-to-noise ratio remains at an acceptable level. This is exemplified in Figure SI-4C, where the derivative has been slightly smoothed to identify the correct maximum. The derivative curve's first peak center accordingly gives the tangent anchor and its height gives the tangent slope. Finally, the tangent's intercept with the data's *baseline*, which is taken as the intensity at 2-5 eV below the cutoff, is determined. We refer to this point as the 'zero-crossing' even though the real baseline often has a small positive y-value, *e.g.*, because of residual gas-phase PE signal. This procedure has no free parameters.

One could in principle define an alternative, but non-standard, approach to determine $E_{cut}$, namely the direct use of the LET curves first inflection point, *i.e.*, the tangent anchor. As noted in the main text, in our high-resolution data, this fixing-point has a 30-60 meV offset from $E_{cut}$ determined via the zero-crossing (this increases to ~150 meV in our lower resolution examples). The use of the tangent anchor may have certain benefits for lower-energy-resolution measurements, where a smeared-out LET has the potential to result in a slanted tangent, potentially yielding too low an $E_{cut}$ value. In such cases, the tangent anchor, being very close to the zero-crossing value in the high-resolution data, has the potential to become a more reliable fixing point, albeit with a small additional error. While we highlight some exploratory resolution-dependent results recorded at a ~125 eV photon energy in Fig SI-4C, the experimental resolving power remains at a factor of 1000, even in the lowest-resolution example. An in-depth study of the behavior of the LET curve and the associated inflection point and tangent results at various and lower resolution settings is warranted, with the goal of exploring the applicability of the aforementioned $E_{cut}$ definitions. However, this is beyond the scope of this work.

We offer some associated words of caution on the alternative methods that have been used to determine cutoff positions and their impact on extracted IEs. While we here adopt a definition of $E_{cut}$ based on the common tangent extrapolation method, some authors may instead report the 'midpoint' of the rise of the cutoff intensity or an alternative point on the LET curve itself, which yields somewhat different IE values. Determining $E_{cut}$ via a fit to the LET curve profile or a traversed intensity level inevitably leads to a dependence of $E_{cut}$ on the shape of the LET curve, which varies with and can even be distorted by the experimental conditions. Adopting an alternative definition of $E_{cut}$ to the tangent approach implemented here creates a discrepancy between the results, positively offsetting the position of $E_{cut}$ and thus all extracted VIEs, in our case amounting to a several 10 to 100 meV shifts. More specifically, as the experimental energy resolution and sharpness of the LET curves decreases, the results obtained using the different cutoff definitions diverge. As noted above, we are currently unable to ascertain whether the lowest-eKE inflection point in the derivative curve or the associated tangent baseline crossing method yields the correct aqueous-phase energetics when lower-energy-resolution conditions are implemented with Method 2. However, we reiterate that this choice has little consequence when we analyze the highest-energy-resolution measurements reported here. We further note that the $E_{cut}$ positions extracted from either inflection points or tangent zero-crossing points in the LET data have been found to be experimentally robust.

## 6 Correction of bias-voltage induced shifts of measured photoelectron kinetic energies

We observed that measured electron kinetic-energy values, $eKE_{meas}$, are distorted for large ($\leq$-30 V) applied bias voltages, at least for the employed high-pressure-tolerant (HiPP) pre-lens HEAs implemented in this work. The distortion manifests itself as a (predominantly linear) increase of $eKE_{meas}$ by a small scaling factor, $s_{distort}$, *i.e.*, if $eKE_{meas}$ is small the effect is negligible but leads to increasingly large errors at larger eKEs. The scaling factor depends on the implemented settings such as the selected pass energy; a systematic study of these technical details has yet to be performed, however. The factor increasingly impacts the accurate determination of VIEs at high photon energies (large spectral widths, $\Delta E_w$), since the cutoff at low eKEs is less affected than the valence features at high eKEs. Conversely, if $\Delta E_w$ is small, the $eKE_{meas}$ values for the cutoff and the valence features are similar, and the error is negligible. Two procedures were employed to determine the scaling factor precisely and correct the resulting VIE values, as explained in the following.

The first procedure relies on a comparison of the energetic position of PE peaks at a known photon energy. By measuring the first and second harmonic spectrum of a specific PE feature, the photon energy can be calculated (see section 3) by applying the equation $E_{ph} = KE_{2nd} - KE_{1st}$. Here, $KE_{2nd}$ is usually much larger than $KE_{1st}$, and if the energy is distorted by an applied bias voltage, then the calculated $E_{ph}$ will be disparate from the real value. This further implies that any photon-energy calibration can be subject to a large error if performed under biased conditions. The same measurement can be repeated in a grounded configuration or $E_{ph}$ can be independently measured, *e.g.*, using gas-phase resonances (again, see section 3), and $s_{distort}$ can then be estimated by comparing the result to the biased case. For example, a gas-phase measurement series defined $E_{ph} = 650.033 \pm 0.014$ eV, while the procedure discussed above yielded $650.216 \pm 0.029$ eV, which gives $s_{distort} = 1.0002842$. An exemplary correction calculation is summarized in Table SI-1. This was the highest distortion identified in our measurements, which in the worst case leads to an offset in the VIE value of up to 350 meV at $E_{ph} = 950$ eV. While the same drawbacks occur as for the high-photon-energy calibration (see section 3), the bias-correction procedure mentioned above is minimally time-consuming and can readily be adopted as an extension of a careful photon-energy calibration procedure.

The second procedure is to measure the desired spectrum with increasing bias-voltage settings and observe the evolution of the resulting VIE value (see Figure SI-6). This yields the voltage onset at which the effect becomes significant and a precise $s_{distort}$ factor can be extracted. If systematically performed for each measurement, a VIE value can be reliably determined free from distortion, either from measurements with a low enough bias voltage or by extrapolation of the behavior. The drawback of this method is that the bias-dependent measurements can be highly time-consuming and the stability of all experimental conditions needs to be precisely maintained during a measurement set. Also, at particularly low bias values, the cutoff may already be affected by either HEA transmission issues or by the analyzer work function, which obscures the real $E_{cut}$ value and may render an accurate extraction of the VIE at lower bias settings difficult or even impossible.

## 7 Discussion of the results of previous studies from Olivieri *et al.*[21] and Ramírez[22]

We first discuss the study of Olivieri *et al.*[21], which reported the work function of water from the LET 'midpoint' as $e\Phi_{water}$ = 4.65 ± 0.09 eV somewhat smaller than our value of $e\Phi_{water}$ = 4.73 ± 0.09 eV. The main text already addresses the fact that the 'midpoint' approach to determine $E_{cut}$ will result in higher $e\Phi_{water}$ values (in our high-spectral-resolution cases by several 10 meV up to ~150 meV), further increasing the offset to our result. The methodological inconsistencies and inaccuracies, which we believe lead to the $e\Phi_{water}$ offsets, are summarized below.

The procedure proposed by Olivieri *et al.*[21] requires a precisely known bias potential to determine a 'correction factor' $c$ or to relate the measured $E_{cut}$ value to the unbiased onset energy of the spectrum. In principle, an associated error can arise from the output of the voltage supply, with another error arising from the fact that the actual potential at the LJ may differ from the applied bias voltage due to additional resistances in the electrical connection. This becomes apparent in the initial analysis of the O 1s shift in Ref. [21], where the voltage is stated to be just 99% translated into the liquid shift (*e.g.*, a 4.00 V bias leads to a shift of 3.96 eV in the liquid water O1s peak); we observed a similar factor of ~98 % (see Figure SI-3). Apparently, this deviation was not corrected for in the determination of the $e\Phi$ difference, and rather the full -40 V bias voltage has been subtracted. This would lead to a rather large error of up to 0.4 eV (40 * 0.99 = 39.6) for all subsequent analyses. On a related note, under typical experimental conditions it can take more than one hour until the equilibration of water (or alternative solution component) adsorption in the spectrometer chamber is established, causing time-dependent changes to the potentials. In the present study, we found that this effect causes energy shifts larger than 100 meV, as shown in Figure SI-2. It is unclear whether or not such behaviors were properly accounted for in the work of Olivieri *et al*. Unfortunately, with their proposed procedure, it is notably impossible to disentangle $e\Phi$ differences from additional extrinsic surface potentials, as both will have the same effect, namely a shift and broadening of the gas-phase O1s peak when ionization occurs close to the liquid surface. Such omission bears the immediate risk that the individual effects add up to a rather large error, manifested here as higher O 1s level ionization energies and lower $e\Phi$ values, with respect to our determinations: 538.21 ± 0.07 eV versus 538.10 ± 0.05 eV and 4.65 ± 0.09 eV versus 4.73 ± 0.09 eV for $VIE_{vac,O1s}$ and $e\Phi_{water}$, respectively. Furthermore, Olivieri *et al.* claim that the thus far published VIE data on aqueous solutions may be incorrect by values as large as the $e\Phi$ difference between the detector and the sample when employing referencing Method 1 (Figure 6). This is clearly not the case for the determination of the lowest ionizing energy of liquid water, associated with the $1b_{1(l)}$ peak position. This VIE should be recorded with sufficiently dilute aqueous solutions, given that the gas-phase referencing method is only (or should only be) applicable after establishing *field-free* conditions, *i.e.*, after compensation of all effects, including any $e\Phi$ difference. While similar arguments could be made for highly concentrated aqueous solutions, shifts in the solvent peaks in those cases do not necessarily originate from $e\Phi$ differences, but may have their origin in genuine changes of the bulk or interfacial aqueous electronic structure.[23]

Next, we briefly comment on the recent study of Ramírez[22] It was already stated in the main text that their values of $VIE_{EF,1b1(l)}$ = 6.94 eV and $e\Phi$ = 4.60 eV reported for a Zobell solution and $VIE_{EF,1b1(l)}$ = 7.06 eV and $e\Phi$ = 4.53 eV reported for a 0.1 M KCl solution (associated errors were not reported), are in disagreement with our results for neat liquid water, *i.e.*, with 50 mM NaI dissolved. $VIE_{vac,1b1(l)}$ values of 11.55 eV and 11.59 eV were also respectively reported for the Zobell and KCl solutions, significantly exceeding the 11.33 ± 0.02 eV value reported here. The Ramírez study notably used the LET 'midpoint' (determined by fitting an EMG function and with its potentially associated flaws, see section 5) to define $e\Phi$ and $VIE_{vac,1b1(l)}$. The corresponding offset of that fixture point with respect to of $E_{cut}$ determined via the tangent method is likely one of the reasons for the offset of Ramírez's[22] values compared to our values. Furthermore, the $1b_1$ peak centroid energy used to determine the distance to $E_{cut}$ was extracted *after* referencing to the Fermi level, which most likely introduced additional offsets to the VIE value, hinging on the correctness of their $VIE_{EF}$ values. Both solutions are unlikely to fully establish streaming-potential free conditions, which would lead to shifts in the $1b_1$-peak to Fermi level distance. For comparison, we would extract a similarly high $VIE_{EF}$ of ~7.0 eV from an aqueous solution with a salt concentration of just a 2.5-5.0 mM, far away from the $I_{str} = 0$ case identified in our work (compare to Figure SI-7). Finally, the same fixed bias voltage of -20.58 V was apparently used to correct all spectra, regardless of solution, despite the fact that some voltage drop inevitably occurs across the liquid (compare Figure SI-3). This voltage drop ultimately depends on the electrical conductivity of the liquid and the distance to the source (which can be assumed to remain constant, however). Using the same fixed value may thus introduce variable errors when applied to spectra of different aqueous solutions.

## Figures

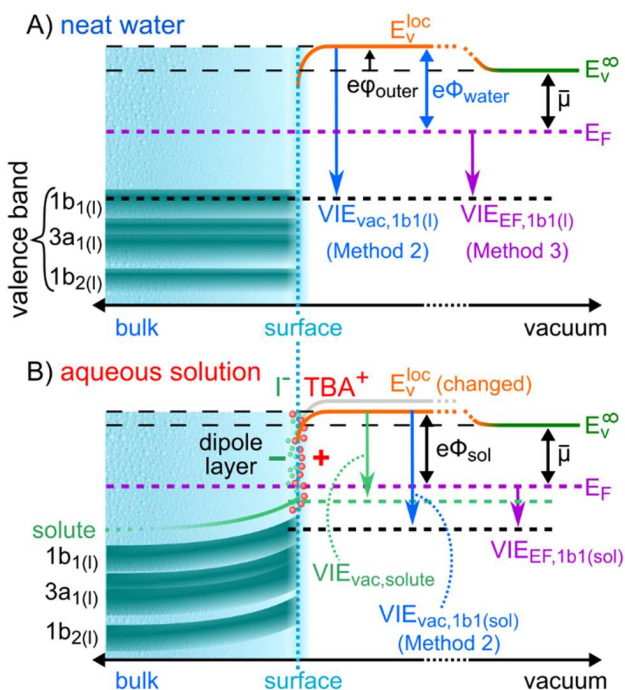

**Figure SI-1:** Schematic energy diagram highlighting the potentials and reference energy levels of relevance to liquid-phase PES experiments, adapted here from Ref. [24] for a neat liquid-water sample (A) and an aqueous solution of TBAI (B). In panel (B), TBA$^+$ and I$^-$ ions accumulate, and change the charge distribution, at the surface layer. Adsorbing strong electron donors or acceptors on semiconducting surfaces induces band bending. The Fermi level within the solution stays fixed (which is termed Fermi-level 'pinning') at its bulk value. PE spectroscopy accesses the interface where the bands are bent, leading to the observed large shift and slight broadening of the TBAI$_{(aq)}$ solution bands.

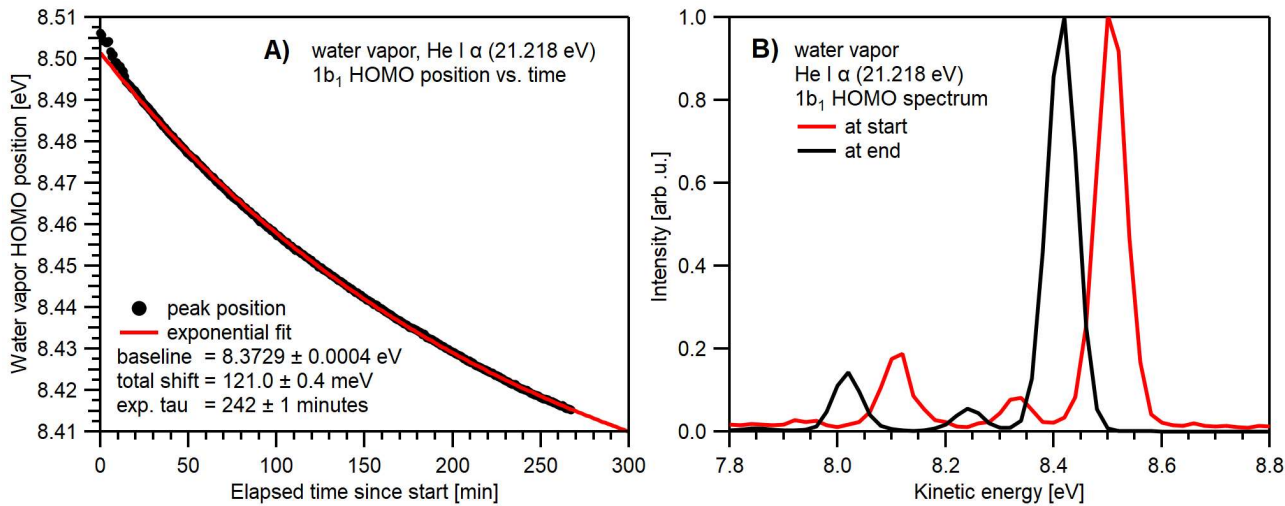

**Figure SI-2:** Water $1b_{1(g)}$ HOMO ionization peak shift after introducing water vapor via a gas nozzle (~$10^{-3}$ mbar of $N_2$-equivalent standard pressure at the sensor, *i.e.*, much higher at the nozzle orifice) into a clean experimental setup. **A)** Position of the $1b_{1(g)}$ vibrational-ground-state ionization peak versus elapsed time after introducing the gas. **B)** PE spectra at the start and the end of this measurement series. The observed shift in this particular example is about 120 meV, but will depend on the initial conditions of the chamber and the absolute gas pressure in the vicinity of the detector entrance. The shift evolves over several hours after introduction of the water vapor; an exact energy reference would only be possible after waiting until the system has equilibrated. Accordingly, it is essential that steady-state, equilibrated conditions are established before any extrinsic perturbing potentials are eliminated or compensated – through electrolyte addition or sample biasing – and associated $\Delta E_{g-l}$ measurements are performed.

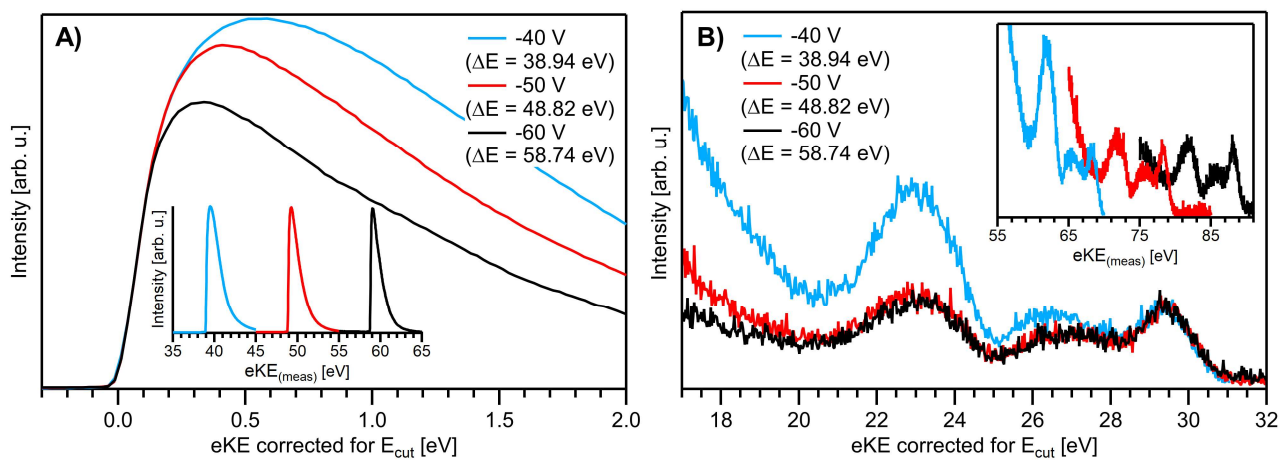

**Figure SI-3:** Demonstration of the rigid energy shift of all liquid features with increasing bias voltage. The liquid water spectrum is measured with He II α emission (40.814 eV) under the influence of -40 V (blue), -50 V (red), and -60 V (black) bias voltages, as directly applied to the LJ with respect to the grounded PES apparatus. **A)** LET and **B)** valence band regions measured as separate spectra to increase the measurement times and signal-to-noise ratio over the spectral regions of interest. In both panels A) and B) the energy shift was compensated after via the tangent method; the LET curves are shown intensity normalized to yield the same tangent slope for better comparability. The bias voltage leads to an energy shift of all liquid spectral features which is compensated

for at each setting to produce an $E_{cut}$ position at zero on the bottom energy axis (the insets show the uncompensated spectra for comparison); both the LET and valence regions are found to shift by the same amount. Note that the sample must be sufficiently conductive to engender this behavior, which is ensured by dissolving a sufficient amount (~50 mM) of salt in an aqueous sample. The bias voltage is translated ~97.5% into an eKE shift in this case; internal resistances and the voltage accuracies of the power supply cause the reduced value. This is, however, of no relevance for the absolute referencing method, Method 2, where knowledge of the voltage or eKE shift values are not required. We observe a change in the LET curve's shape towards higher eKE's (panel A) under the employed conditions, which however is inconsequential to the shape and inflection point of the cutoff feature. Specifically, the point that determines the cutoff energy (the baseline crossing point of the tangent defined by the spectral derivative; see Figure 1 and SI-4) leads to an almost identical $\Delta E_W$ value within our error bars, *i.e.*, our method is invariant to the particular LET shape beyond the initial rise.

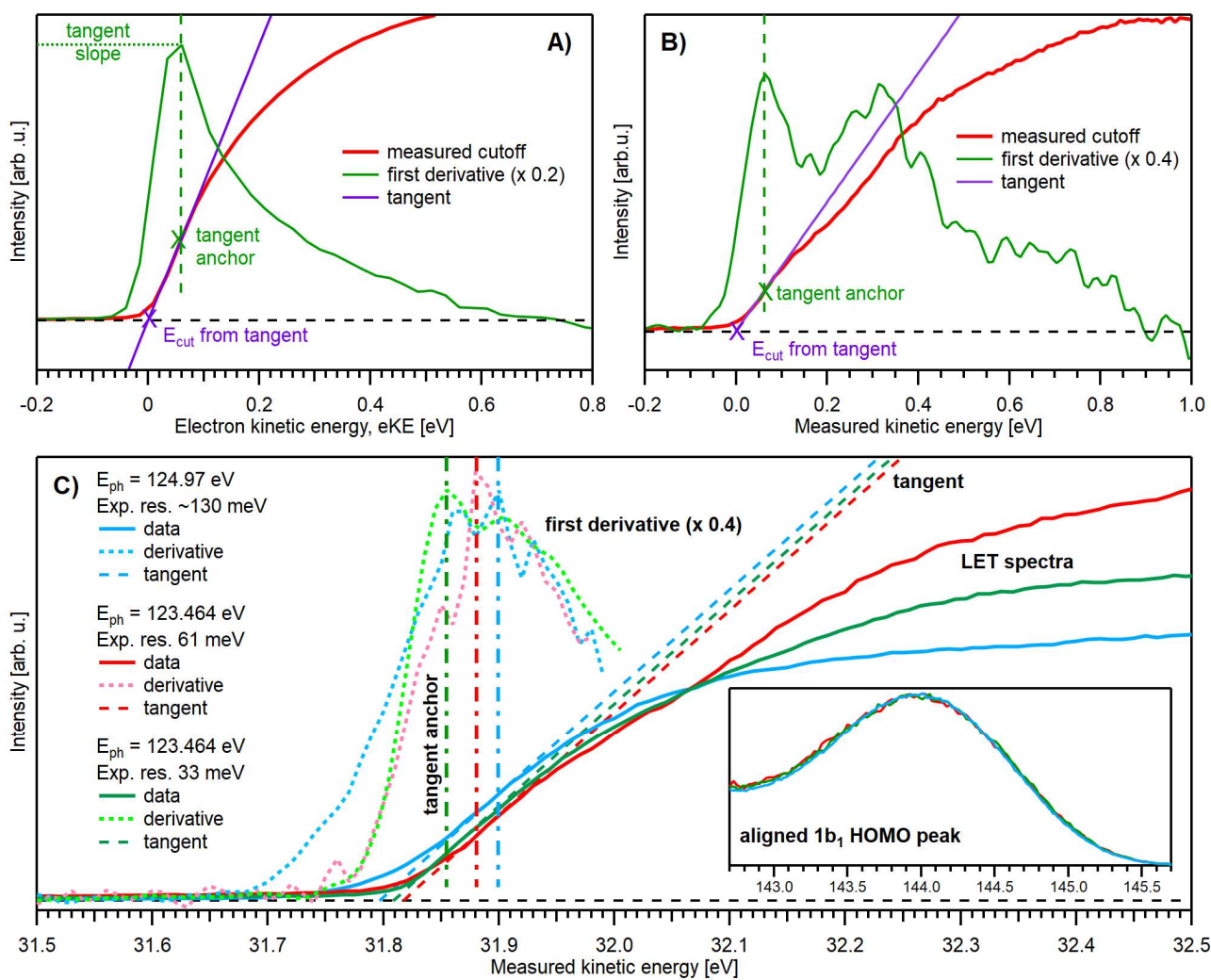

**Figure SI-4:** Procedure for extracting the LET cutoff position, $E_{cut}$, via the tangent method exemplified using PE spectra measured with He IIα emission (40.814 eV) in panel A and synchrotron radiation (~123.5-125 eV) in panel B; the bias voltage was -20 V in panel A and -32 V in panels B and C. **A)** The first derivative (green) of the measured data (red) is calculated. The derivative's first maximum (= inflection point with maximum change in slope in the data) automatically determines the tangent anchor point and slope; no free parameters

exist in such a determination. The position of intersection with the baseline (= signal intensity at lower energies below the LET, if non-zero), which determines $E_{cut}$, is calculated from the tangent equation. $E_{cut}$ defines the energy scale of the photoelectron after leaving the liquid as eKE = 0. **B)** exemplifies data with slight intensity variations in the LET curve, yielding multiple maxima in the derivative. Irrespective of such shape variations, the same procedure as in A), *i.e.*, using the derivative's first maximum, gives the correct $E_{cut}$. Here, the derivative was additionally smoothed using a 3-point binominal smoothing to reduce noise. **C)** LET spectra, derivative and determined tangents from different beamtimes, spanning 8 months, and with different experimental resolutions of approx. 130 meV (blue), 61 meV (red), and 33 meV (green). Here, the spectra were aligned in a way which maximizes overlap of the $1b_1$ HOMO peak in the associated valence band spectra (see inset), where the different photon energy for the blue spectra was taken into account; this accounts for differences in the applied bias value and makes it possible to directly compare the LET spectra. Furthermore, the spectra were scaled to give an equal slope for all tangents, which makes it easy to see slight shifts in the determined positions. It can be seen that the tangent method is exceptionally robust, with $E_{cut}$ values being stable within 20-30 meV. While the tangent anchor is also very stable within ~50 meV, it is apparent that the tangent zero-crossings and tangent anchor points diverge, with the latter shifting slightly towards higher eKEs, as the LET spectrum is broadened by the decreasing resolution.

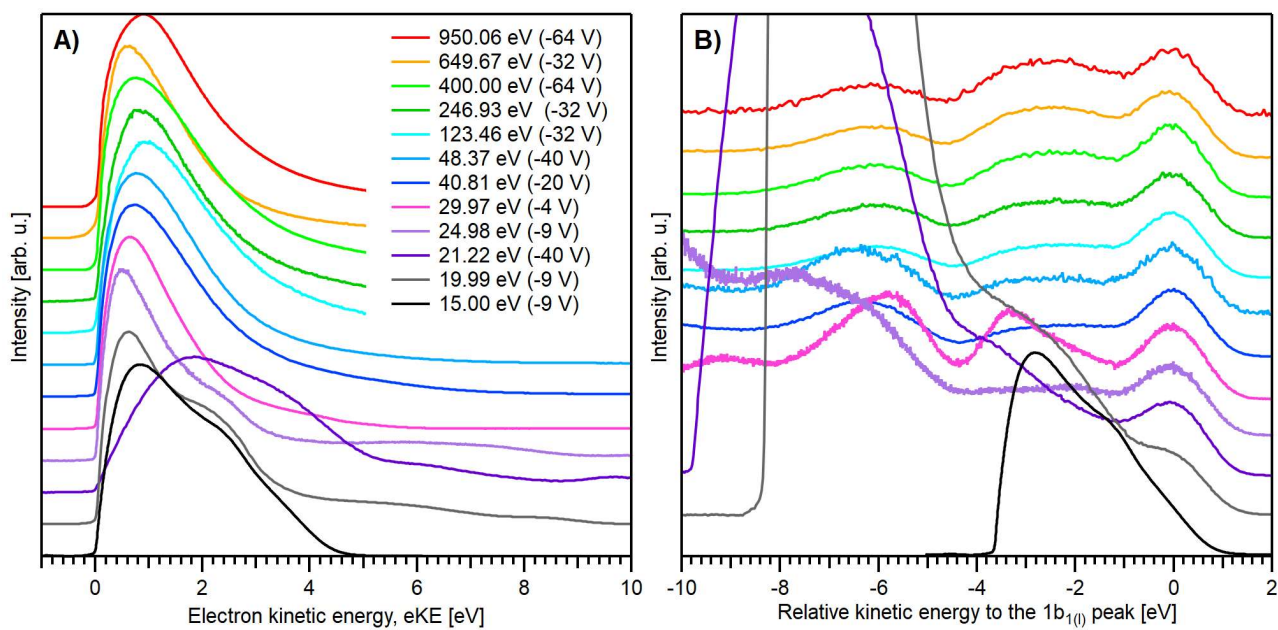

**Figure SI-5:** Representative PE spectra measured at photon energies spanning 15 eV to 950 eV and analyzed in this study. The small residual gas-phase signal in the cutoff region has been subtracted for the He lamp spectra (21.22 eV, 40.81 eV and 48.37 eV); the gas-phase contribution for all other spectra was negligible. **A)** The LET spectra scaled to yield approximately the same height at the cutoff region, where the spectral cutoff has been aligned to zero eKE. **B)** Spectra measured at the same photon energies as in A) and energetically aligned to the $1b_1$ HOMO ionization peak and vertically scaled to yield approximately the same associated peak height. Valence spectra for the higher photon energies (>48 V) were measured separately from the cutoff spectra under identical experimental conditions and over a limited spectral range (*i.e.,* in different but sequential data acquisitions to the cutoff spectra and making sure to adopt the same HEA pass energy and lens table). This allowed us to maximize acquisition times over the spectral regions of interest and optimize the signal-to-noise ratios.

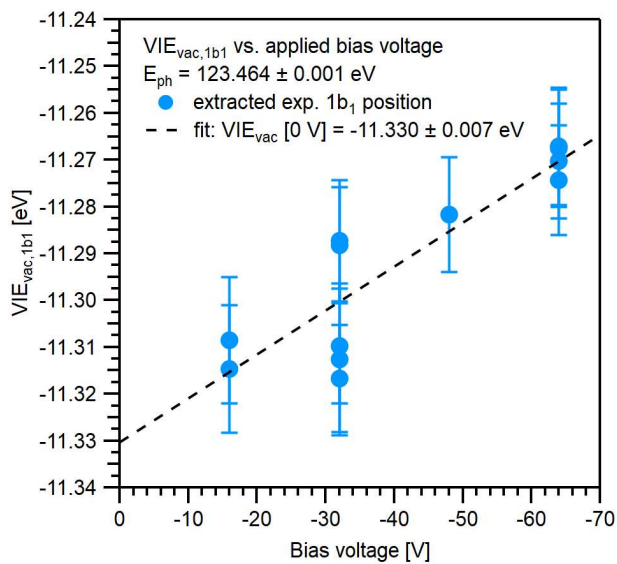

**Figure SI-6:** Extracted $VIE_{vac,1b1(l)}$ value from a series of measurements over two days with increasing and decreasing applied bias voltage (both directions were probed) as blue dots. The dashed black line is the linear regression of all data points which gives the extrapolated VIE value for 0 V bias voltage. The VIE seemingly decreases with lower bias voltage, which indicates an affected energy scale of the HEA. This deviation gets more severe with higher photon energies, since the energy distance between the cutoff and the valence band features increases, which in turn increases the relative error in the measured eKEs. In cases where only -64 V measurements were performed, the VIE value was corrected by the procedure described in SI section 6.

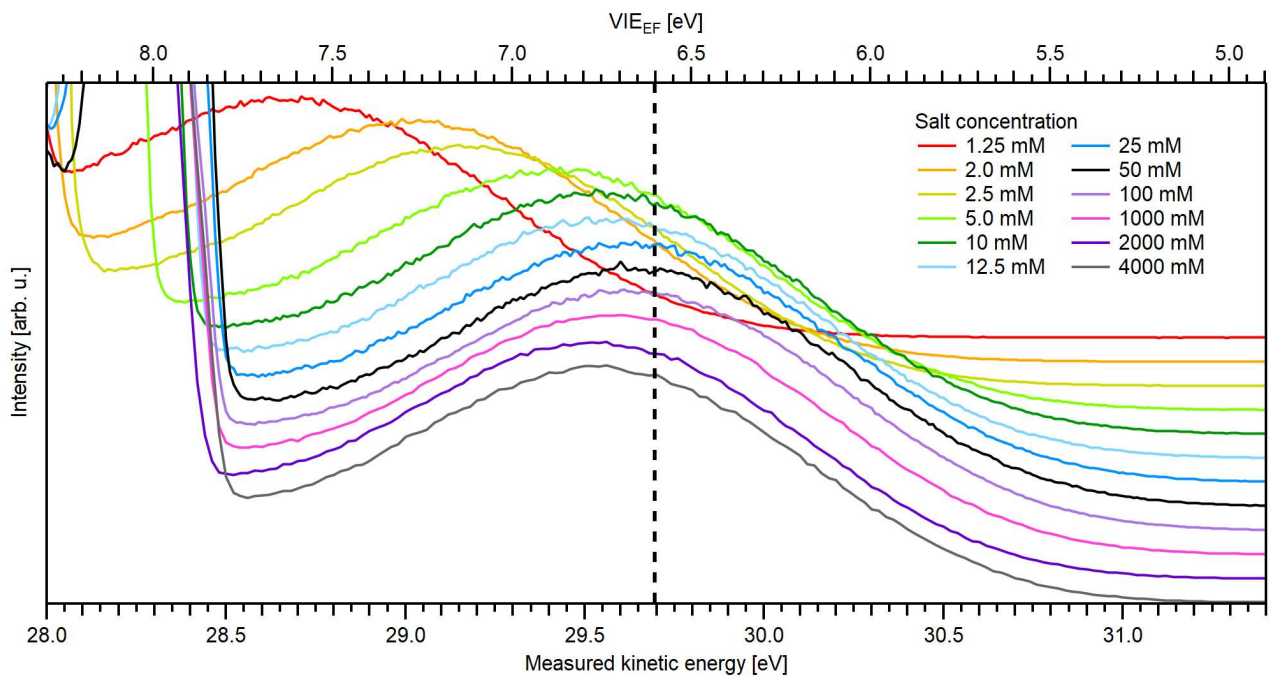

**Figure SI-7:** Demonstration of the introduced energetic shifts of the gas- and liquid-phase water PE features for different salt concentrations; NaI and NaCl dissolution gives similar results. The liquid spectra were measured with He II α (40.813 eV) radiation and a grounded LJ. A metallic reference PE spectrum defines the top $VIE_{EF}$

energy scale; the position of the metal reference sample's Fermi edge was unchanged, irrespective of the presence of flowing solutions of different concentrations. It is apparent that the liquid $1b_1$ peak energy shifts widely with respect to the Fermi edge, with a minimum $VIE_{EF,1b1(l)}$ value reached upon dissolution of salt to 25-50 mM concentrations. Higher and lower concentrations shift all liquid PE features towards lower eKE values, which is interpreted as the effect of an uncompensated streaming potential. The optimal concentration of 25-50 mM needed to suppress the streaming current, $I_{str}$, is in good agreement with previously reported values.[7, 8] At very low concentrations (<1 mM), the sample would be further positively charged due to its insufficient conductivity and means to compensate photoionization-induced electron loss, leading to further retardation of the emitted photoelectrons. Without considering such extrinsic potentials, an arbitrary $VIE_{EF,1b1}$ value would be determined.

## Tables

| hν (eV) | measured $E_{cut}$ (eV) | measured $eKE_{1b1}$ (eV) | measured $VIE_{vac,1b1(l)}$ (eV) | corrected $E_{cut}$ (eV) | corrected $eKE_{1b1}$ (eV) | corrected $VIE_{vac,1b1(l)}$ (eV) |
|---|---|---|---|---|---|---|
| 249.992 ± 0.020 | 62.368 | 301.145 | -11.23 ± 0.03 | 62.350 | 301.059 | **-11.28 ± 0.04** |
| 400.007 ± 0.030 | 62.374 | 451.182 | -11.19 ± 0.03 | 62.357 | 451.054 | **-11.31 ± 0.04** |
| 650.033 ± 0.030 | 62.242 | 701.184 | -11.08 ± 0.05 | 62.224 | 700.985 | **-11.27 ± 0.05** |
| 950.063 ± 0.030 | 62.252 | 1001.249 | -10.94 ± 0.05 | 62.234 | 1000.965 | **-11.33 ± 0.08** |

**Table SI-1:** Exemplary comparison of VIE values from a measurement with a bias of -64 V before and after correction with $s_{distort}$ = 1.0002842. The correction factor was applied to both $E_{cut}$ and the valence-band PE spectrum, which then yields the corrected VIE values.